\documentclass[fleqn,onecolumn,useAMS]{mnras}

\usepackage{graphicx}	
\usepackage{amsmath}	
\usepackage{amssymb}	
\usepackage{multicol}        
\usepackage{bm}		
\usepackage{pdflscape}	

\newcommand{\etal}{et al. }
\newcommand{\adhoc}{{\it ad hoc} }
\newcommand{\adhocp}{{\it ad hoc}}
\newcommand{\mytorus}{MYTORUS }
\newcommand{\mytorusp}{MYTORUS}

\newcommand{\rxte}{{\it RXTE} }

\newcommand{\asca}{{\it ASCA} }
\newcommand{\ascap}{{\it ASCA}}

\newcommand{\xmm}{{\it XMM-Newton} }
\newcommand{\xmmp}{{\it XMM-Newton}}
\newcommand{\chandra}{{\it Chandra} }

\newcommand{\suzaku}{{\it Suzaku} }

\newcommand{\bsax}{{\it BeppoSAX} }

\newcommand{\integral}{{\it INTEGRAL} }

\newcommand{\swift}{{\it Swift} }

\newcommand{\nustar}{{\it NuSTAR} }
\newcommand{\nustarp}{{\it NuSTAR}}
\newcommand{\nicer}{{\it NICER} }
\newcommand{\nicerp}{{\it NICER}}
\newcommand{\xrism}{{\it XRISM} }

\newcommand{\apec}{{\sc APEC} }
\newcommand{\pexrav}{PEXRAV }
\newcommand{\pexravp}{PEXRAV}
\newcommand{\pexmon}{PEXMON }
\newcommand{\pexmonp}{PEXMON}
\newcommand{\rdblur}{RDBLUR }
\newcommand{\rdblurp}{RDBLUR}
\newcommand{\reflionx}{REFLIONX }

\newcommand{\chisq}{$\chi^{2}$ }

\newcommand{\olyap}{O~{\sc viii}~Ly$\alpha$}

\newcommand{\nelyap}{Ne~{\sc viii}~Ly$\alpha$}
\newcommand{\arlya}{Ar~{\sc xviii}~Ly$\alpha$ }

\newcommand{\fekalfa}{{Fe~K$\alpha$} }
\newcommand{\fekalfap}{{Fe~K$\alpha$}}

\newcommand{\fekbeta}{{Fe~K$\beta$} }
\newcommand{\fekbetap}{{Fe~K$\beta$}}
\newcommand{\nika}{{Ni~K$\alpha$} }

\newcommand{\fexxvres}{Fe~{\sc xxv}(r) }
\newcommand{\fexxvresp}{Fe~{\sc xxv}(r)}

\newcommand{\fexxvip}{Fe~{\sc xxvi}}
\newcommand{\feklya}{{Fe~{\sc xxvi}~Ly$\alpha$} }

\newcommand{\src}{NGC~4388 }
\newcommand{\srcp}{NGC~4388}

\newcommand{\nhs}{{$N_{\rm H,S}$ }}
\newcommand{\nhsp}{{$N_{\rm H,S}$}}
\newcommand{\nhz}{{$N_{\rm H,Z}$ }}
\newcommand{\nhzp}{{$N_{\rm H,Z}$}}
\newcommand{\cxispinp}{$C_{\rm PIN:XIS}$}
\newcommand{\cxispin}{$C_{\rm PIN:XIS}$ }

\newcommand{\logxip}{$\log{\xi}$}

\newcommand{\thetaobs}{{$\theta_{\rm obs}$} }

\newcommand{\relxill}{{\sc RELXILL} }
\newcommand{\relxillp}{{\sc RELXILL}}

\newcommand{\tablectratesp}{{Table~1}}
\newcommand{\tablespfits}{{Table~2} } 
\newcommand{\tablespfitsp}{{Table~2}}

\newcommand{\tablefluxesp}{{Table~3}}
\newcommand{\tablerelxillfits}{{Table~4} }
\newcommand{\tablerelxillfitsp}{{Table~4}}
\newcommand{\tablehistory}{{Table~A1} }
\newcommand{\tablehistoryp}{{Table~A1}}

\usepackage[T1]{fontenc}
\usepackage{ae,aecompl}

\usepackage{newtxtext,newtxmath}

\title[NGC 4388: A Test Case for Relativistic Disk Reflection and Fe K Fluorescence Features]{NGC 4388: A Test Case for Relativistic Disk Reflection and Fe K Fluorescence Features}
\author[T. Yaqoob]{Tahir Yaqoob$^{1,2}$, P. Tzanavaris$^{1,2,3}$, S. LaMassa$^{4}$ \\
	$^1$Center for Space Science and Technology, University of Maryland, Baltimore County, 1000 Hilltop Circle, Baltimore, MD 21250, USA \\
  $^2$Center for Research and Exploration in Space Science and Technology, NASA/GSFC, Mail Code 662, Greenbelt, MD 20771, USA \\
	$^3$The American Physical Society, Hauppauge, New York 11788,USA \\  
	$^4$Space Telescope Science Institute, 3700 San Martin Drive, Baltimore, MD 21218, USA \\
}

\date{Received 2023}

\pubyear{2023}

\begin{document}
\label{firstpage}
\pagerange{\pageref{firstpage}--\pageref{lastpage}}
\maketitle

\begin{abstract} 

We present a new analysis of the \suzaku X-ray spectrum of the
Compton-thin Seyfert 2 galaxy \srcp.  The spectrum above $\sim2$~keV
can be described by a remarkably simple and rather mundane model,
consisting of a uniform, neutral spherical distribution of matter,
with a radial column density of $2.58\pm0.02 \times 10^{23} \ \rm
cm^{-2}$, and an Fe abundance of $1.102^{+0.024}_{-0.021}$ relative to
solar. The model does not require any phenomenological adjustments to
self-consistently account for the low-energy extinction, the \fekalfa
and \fekbeta fluorescent emission lines, the Fe~K edge, and the
Compton-scattered continuum from the obscuring material. The spherical
geometry is not a unique description, however, and the
self-consistent, solar abundance \mytorus model, applied with toroidal
and non-toroidal geometries, gives equally good descriptions of the
data. In all cases, the key features of the spectrum are so tightly
locked together that for a wide range of parameters, a relativistic
disk-reflection component contributes no more than $\sim2\%$ to the
net spectrum in the 2--20~keV band.  We show that the commonly invoked
explanations for weak X-ray reflection features, namely a truncated
and/or very highly ionized disk, do not work for \srcp.  If
relativistically-broadened \fekalfa lines and reflection are
ubiquitous in Seyfert 1 galaxies, they should also be ubiquitous in
Compton-thin Seyfert 2 galaxies. The case of \src shows the need for
similar studies of more Compton-thin AGN to ascertain whether this is
true.

\end{abstract}



\section{Introduction}
\label{intro}

It is nearly three decades since the first reports of
relativistically-broadened \fekalfa line emission in the X-ray spectra
of active galactic nuclei (AGNs) with moderate spectral-resolution CCD
detectors (Tanaka \etal 1995; Yaqoob \etal 1995).  Since then, studies
using the broad \fekalfa lines, and associated reflection spectra, to
probe the properties of the inner accretion disk and angular momentum
(spin) of the central black-hole in AGN, have proliferated enormously,
transforming the literature on AGN X-ray spectroscopy in a generation
(e.g., see Reynolds 2019, 2021, and references therein). Originally,
studies focussed on type~1, unobscured AGN because the X-ray spectra
of obscured AGN are more complex, exhibiting line-of-sight extinction,
as well as narrow \fekalfa line emission and Compton-reflection
continuum emission from nonrelativistic, distant matter. However, as
modelling techniques for deconvolving the relativistic and
nonrelativistic components improved, obscured AGN have been
increasingly included alongside unobscured AGN for probing the inner
accretion disk and black-hole spin (e.g., Tzanavaris \etal 2021, and
references therein).

Over the years, relativistic disk-reflection models have become
increasingly sophisticated, involving over a dozen parameters (e.g.,
see Dauser \etal 2013, 2022, and references therein), and it has now
become common practice to {\it begin} modelling an AGN X-ray spectrum
with the most complex scenarios, regardless of whether the data
require such complexity or not.  In particular, spectral-fitting with
complex, over-parameterized models proceeds without testing whether a
relativistically-broadened \fekalfa line emission can actually be
detected independently of the other features in the spectrum. There
are so many parameters and freedom in the state-of-the-art
disk-reflection models, that it is almost impossible that they will
not fit the data. In the present paper we show results for an
important case study of an AGN that should exhibit strong reflection
and relativstically-broadened Fe fluorescence features according to
standard models and expectations, but which in fact has an X-ray
spectrum that has a very simple and mundane interpretation. The
general principle that the case study highlights is that, in order to
truly understand AGN X-ray spectra we must roll back the approach of
immediately fitting the data with the most complex models available,
and revert back to rigorously allowing the data to guide the
model-fitting, as opposed to letting the models guide the analysis.

Many of the works on \suzaku observations of AGN were published before
the existence of models that were able to self-consistently account
for the \fekalfa line emission and Compton-reflection from distant,
nonrelativistic matter (e.g., see Nandra \etal 2007, and references
therein). Such models enable disentanglement of the relativistic and
distant-matter components. In the light of parallel improvements in
the relativistic disk-reflection models since the \suzaku era, it is
imperative to revisit some of the \suzaku observations to apply the
newer methodologies.  With this motivation, we have found that the
\suzaku X-ray spectrum of the Seyfert~2 galaxy \src is a very
important test case.  The original analysis of the \suzaku data
presented by Shirai \etal (2008) fitted a host of heterogenous, \adhoc
models that included basic disk-reflection models that were available
at the time, in combination with empirical model components.  As with
the analysis of many other AGN X-ray spectra at that time, there was
no overall insight into a robust interpretation of the \suzaku X-ray
spectrum of \srcp, given the approach of fitting \adhoc models that
were unable to treat the physics self-consistently.  The results of
other historical studies of the same \suzaku spectrum of \src are
summarized in the appendix.  We studied the same \suzaku spectral data
of \srcp, and found that above $\sim2$~keV, the spectrum can be very
simply explained by extinction and reflection in a Compton-thin
($N_{\rm H}<1.25 \times 10^{24} \ \rm cm^{-2}$), uniform, neutral,
solar-abundance, spherical distribution of matter, with no requirement
for any relativistic disk emission component. Although the uniform
spherical model is not a unique interpretation of the data, it is rare
to be able to account for the \fekalfa line and high-energy continuum
in an AGN X-ray spectrum with such a simple model. This makes \src an
important reference source that generalized models of AGN involving
disk-reflection spectra have to explain. For example, relativistic
disk-reflection models fitted to AGN often require a factor of several
overabundance in Fe (e.g., Garc\'{i}a \etal 2018, and references
therein), yet our analysis of the \suzaku spectrum of \src tightly
constrains the Fe abundance to within $10\%$ of solar.  Our study
enforces the condition that the Fe abundance in the disk and in
distant matter cannot be different in the same astrophysical
source. We are not aware of any similar study that enforces such a key
condition.  In the present paper we systematically derive upper limits
to the contribution of a relativistic disk component contribution to
the \suzaku spectrum of \srcp, for a wide range of parameter space,
and find the contribution to be negligible.

The study in the present paper is restricted to the \src \suzaku
observation and focuses on the high-energy features in the spectrum
above $\sim2$~keV, although spectral-fitting analysis is performed
down to $\sim0.6$~keV.  \src is one of the brightest AGN in the X-ray
band, so it has been observed by every X-ray astronomy mission since
\ascap.  The \suzaku spectrum still remains the best for
simultaneously studying the \fekalfa line, continuum absorption, and
the Compton-reflected continuum. The latter requires coverage beyond
10~keV, out to at least 20~keV, and this condition is not satisfied by
observations of \src with \asca (Iwasawa \etal 2007), \chandra (Shu
\etal 2011; Yi \etal 2021, and references therein), \xmm (Elvis \etal
2004; Beckmann \etal 2004), or {\it NICER} (Miller \etal 2019).
Although \bsax (Risaliti 2002) and \nustar (Masini \etal 2016; Kamraj
\etal 2017; Ursini \etal 2019) do have the high-energy coverage, they
have poor spectral resolution in the \fekalfa emission-line and Fe-K
region of the X-ray spectrum. In the present work, more detailed
comparisons of our results from the \suzaku spectral analysis of \src
with results from previous studies of the same data and from
observations with other X-ray astronomy missions can be found in the
appendix.  Here, we note only that the model-fitting of the spectra
from other missions is very heterogeneous, and that none of the
historical studies attempted to explain the high-energy X-ray spectra
with a simple, uniform, spherical X-ray reprocessor.

The remainder of the present paper is organized as follows. In
\S\ref{obsdata} we describe the data reduction procedures for
constructing the spectra from the different \suzaku instruments, and
for preparing the instrument spectral responses. In \S\ref{spfitting},
we describe the spectral-fitting analysis, and present the results for
spherical and other models of the distant matter distribution. In
\S\ref{relline} we present the method and results for obtaining upper
limits on the contribution of relativistic disk emission to the
\suzaku X-ray spectrum of \srcp.  Finally, in \S\ref{conclusions}, we
give a summary of results, and our conclusions. Relevant results from
historical studies of \src with different missions are summarized in
the appendix.

\section{Data Reduction}
\label{obsdata}

The \suzaku data were reduced and analyzed in exactly the same way as
described in Yaqoob \etal (2015) for Mkn~3 (including spectral
extraction details and backgroud subtraction methods). As for the case
of Mkn~3, the GSO spectrum quality for \src was too poor and
unreliable to use for the detailed spectral study that is the goal of
the present paper, so we rejected the GSO spectrum. The present study
focuses on spectral data from the XIS CCD detector (Koyama \etal~2007)
and the HXD PIN high-energy detector (Takahashi \etal~2007).
Excluding lower and upper energy ranges that gave negative counts in
the background-subtracted spectra, and the XIS range 2.1--2.45~keV due
to known, large calibration systematics around the mirror Au M edges,
resulted in the final energy ranges shown in \tablectratesp, which
also shows the background-subtracted count rates in these energy
ranges.  We note that the background-subtracted count rates for XIS
and HXD/PIN are 85.9\% and 35.8\% respectively, of the net on-source
count rates.

The absolute flux calibrations of the PIN and XIS have a systematic
difference, and we denote the PIN:XIS corss-normalization ratio by
\cxispinp. The \src observation was done in the so-called
``XIS-nominal'' pointing mode, and for this configuration, the
recommended cross-normalization factor is 1.16 (see Yaqoob 2012,
including a fuller discussion, covering situations that warrant
deviations from this value). We fixed \cxispin at the recommended
value for all spectral-fitting analyses.

\begin{table}
\caption[Exposure Times and Count Rates for the NGC~4388 \suzaku Spectra]
{Exposure Times and Count Rates for the \suzaku Spectra}
\begin{center}
\begin{tabular}{lllc}
\hline

Detector & Exposure & Energy Range & Rate$^{a}$ \\
	 &  (ks)   & (keV)    &  (ct/s) \\
\hline
XIS & 112.4 & 0.6--2.1, 2.45--9.0 & $0.3984 \pm 0.0011$ \\
PIN & 104.4$^{b}$ & 14.0--40.8 & $0.217\pm0.025$ \\
\hline
\end{tabular}
\end{center}
$^{a}$ Background-subtracted count rate in the energy bands specified. $^{b}$ After dead-time correction.
\end{table}

\section{Spectral Fitting Analysis}
\label{spfitting}

One goal for the spectral-fitting analysis for \src is to constrain
both the line-of-sight and global column densities of matter
surrounding the X-ray source, by modelling the \fekalfa line emission
self-consistently with respect to the X-ray absorption and
Compton-scattered continuum (reflection) due to the same matter
distribution.  Such self-consistent modelling of the distant-matter
features will reduce the degeneracy encountered with \adhoc models,
when assessing the possible spectral contributions from features
originating in matter that is closer to the putative black hole, such
as relativistically broadened \fekalfa line emission and reflection
from an accretion disk (see \S\ref{relline}).  In \S\ref{sphericalfit}
we describe the results of spectral fitting with a uniform spherical
model of circumnuclear matter, and in \S\ref{mytorusfits} we give the
results of fitting the \mytorus model (Murphy \& Yaqoob 2009).  The
latter model will also be used in a mode that is not restricted to a
toroidal geometry.  Since the \mytorus spectral-fitting model was
released, many other torus models have become available (Ikeda, Awaki,
and Terashima 2009; Brightman and Nandra 2011; Liu and Li 2014; Furui
\etal 2016; Balokovi\'{c} M. 2018; Buhcner \etal 2019; Tanimoto \etal
2019).  Descriptions and comparisons of different torus models can be
found in Saha, Markowitz, and Buchner (2021), who conclude that, in
general, current data are unable to distinguish between them.

We used XSPEC (Arnaud 1996)
v12.11.1 \footnote{http://heasarc.gsfc.nasa.gov/docs/xanadu/xspec/}
for spectral fitting.  Galactic absorption with a column density of
$2.60 \times 10^{20} \ \rm cm^{-2}$ (Heiles and Cleary 1979) was
included in all of the models The same distant-matter models were used
in the study of Mkn~3 (Yaqoob \etal 2015), and details of the various
assumptions and caveats can be found there.  In order to account for
possible calibrations systematics in the energy scale, and/or possible
very mild ionization of the distant matter, we allowed the redshift
parameter of the reprocessor models ($z_{\rm R}$), to vary relative to
the cosmological redshift of $z=0.008419$ (Lu \etal 1993).  The
\suzaku CCD spectra can detect shifts in the \fekalfa line peak that
are as small as $\sim10$~eV. We note that a precision \chandra HETG
measurement of the centroid energy of the \fekalfa line in \src of
$6.393\pm0.004$~keV (Shu \etal 2011) constrains the line emitter to be
essentially neutral.

The XSPEC convoluton model GSMOOTH is used to apply velocity
broadening to the intrinsic line widths. Details can be found in
Yaqoob \etal (2015), including the relation between the model width
parameter, $\sigma_{0}$ (keV), and the FWHM ($\rm km \ s^{-1}$).  In
all of the models, the \fekalfa line intensity and equivalent width
(EW) are not free parameters because the line is produced
self-consistently in the models. The models have no free parameters at
all for the \fekbeta line because the line is calculated
self-consistently with respect to the \fekalfa line. Both the \fekalfa
and \fekbeta lines have an associated Compton shoulder, and each of
these components have no free parmeters because they are again
calculated self-consistently with respect to the entire model.

The observed continuum fluxes, $F_{\rm obs}$, and the observed
(rest-frame) luminosities, $L_{\rm obs}$, will be given in various
energy bands. The intrinsic continuum luminosities, $L_{\rm intr}$,
were calculated by turning off all absorption and scattering
components.  For calculating luminosities, we use a standard, flat
$\Lambda \rm CDM$ cosmology, adopting $H_{0} = 73.3 \ \rm km \ s^{-1}
\ Mpc^{-1}$, and $\Omega_{\Lambda} = 0.70$ (Wong \etal 2020).  All
astrophysical model parameter values will be given in the rest frame
of \srcp, unless otherwise stated.

\begin{figure}
\centerline{
        \includegraphics[height=11cm,angle=270]{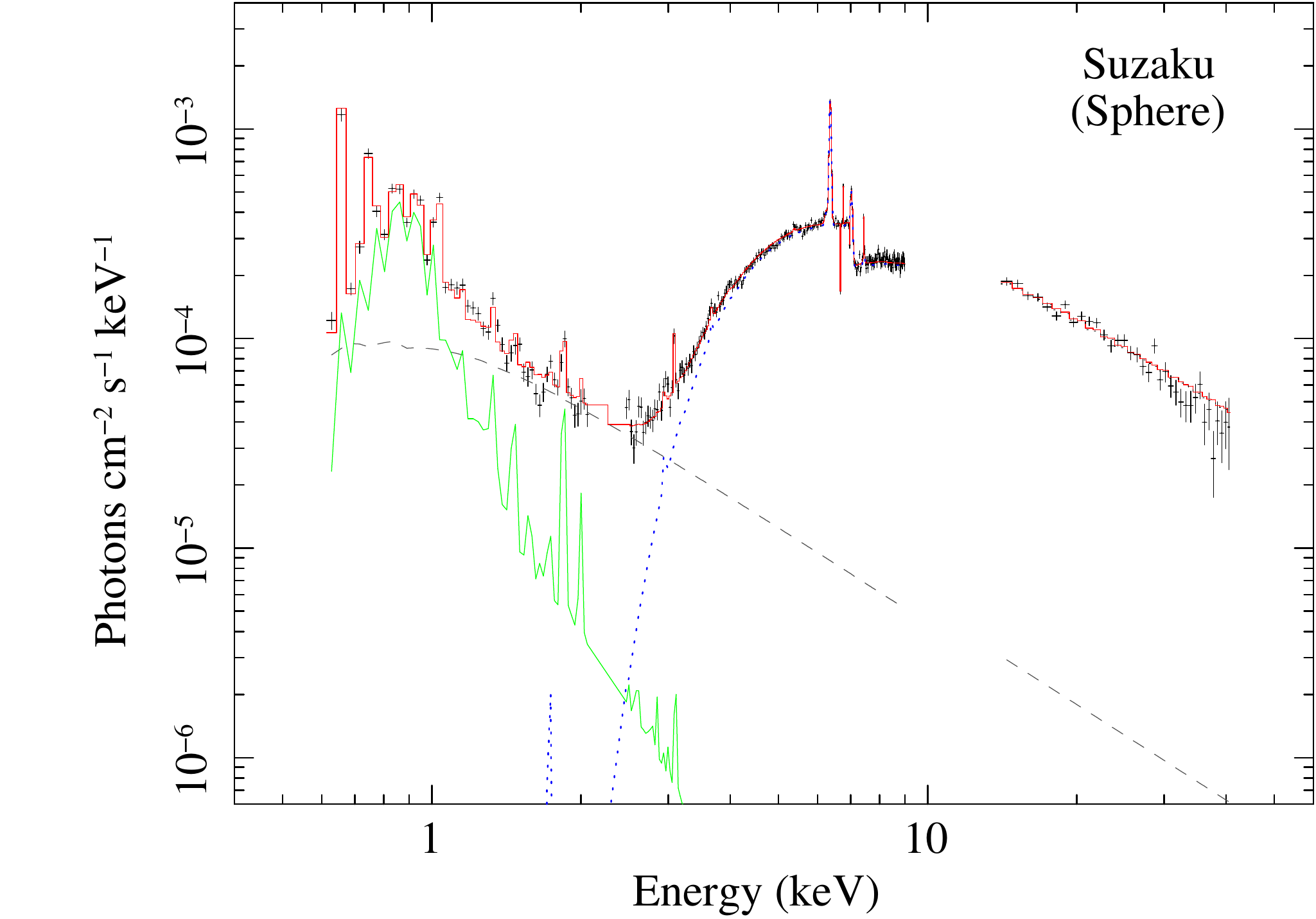}
        }
\caption{\small (a) Full, broadband spectral fit to the
suzaku \src data with the best-fitting, uniform, neutral,
spherical X-ray reprocessing model (see
\S\ref{sphericalfit}, \tablespfitsp). Shown are the unfolded
pectrum (black), the net model spectrum (red) the
ontribution from the spherical X-ray reprocessor (dotted
lue curve), the second power-law continuum (grey dashed
line), and the optically-thin thermal emission ({\sc apec})
component (green). As noted in Yaqoob \etal (2015), ``in an
unfolded spectrum, some emission-line features appear to be
artificially narrower than they really are''.  The same fit
overlaid instead on the counts spectrum in
Fig.~\ref{fig:szsphdatrat} does not suffer from this
effect. Fig.~\ref{fig:szsphdatrat} also shows the data/model ratios
for this fit.}
\label{fig:szsphuspec}
\end{figure} 

\subsection{Spectral Fits with a Uniform Spherical X-ray Reprocessor Model}
\label{sphericalfit}

In this section we model the circumnuclear matter in \src with a very
simple model consisting of a fully-covering uniform, neutral,
spherical distribution of matter centered on the X-ray source, in
which fluorescent line emission and the Compton-reflection continuum
are calculated self-consistently. This is the simplest possible
configuration of the global distribution of the X-ray reprocessor, and
we use the model of Brightman and Nandra, 2011 (hereafter
BN11). Historical observations of \src going back over two decades
show that energy of the prominent \fekalfa emission line indicates
that it originates in neutral matter (see appendix). The application
of the BN11 model to \src follows that for Mkn~3 (Yaqoob \etal 2015),
and full details, its parameters, and various caveats, can be found in
that study.

In addition to the BN11 model itself, the spectral data require other
model components, none of which are unusual, as similar components are
found in other AGN in general. One of these components is the second
power-law continuum mentioned earlier, characterized by $f_{s}$, the
fraction of the direct continuum that is scattered into the line of
sight by optically-thin matter, extended on the same or larger size
scale as the main X-ray reprocessor.  Another spectral component
included is thermal emission from optically-thin matter, and this is
modelled using the XSPEC model {\sc apec}. In addition, we found it
necessary to include spectral components for eight additional (narrow)
emission or absorption features that were not accounted for by any of
the other model components. For these, we used eight simple Gaussian
components. Finally, a uniform, neutral, absorbing screen with column
density $N_{H,1}$ (a free parameter), was applied to all of the
spectral components.  This column density may correspond to absorption
in the host galaxy of \src itself, and/or a small amount of additional
absorption situated further out than the circumnuclear region modelled
by the BN11 component. For the sake of reproducibility, we give the
exact XSPEC model expression used to set up the model:

\begin{eqnarray}
\rm
 model \  = \  constant<1>*phabs<2>zphabs<3>( & + & \nonumber \\ 
\rm gsmooth<4>*(atable\{sphere0708.fits\}<5>) & + & \nonumber \\
\rm constant<6>*zpowerlw<7>  + \rm  apec<8> & + & \nonumber \\
\rm \sum_{n=9}^{n=17} { zgauss<n> } ) 
\end{eqnarray}

In the above expression, we identify $\rm constant<1>=C_{\rm
  PIN:XIS}$, $\rm phabs<2>=$ Galactic column density, $\rm zphabs<3>=
N_{H,1}$, and ${\rm constant}<6> = f_{s}$ (associated with the distant
scattering continuum).  The $\rm apec<8>$ and $\rm zgauss<n>$
components refer to the optically-thin thermal continuum emission and
the narrow Gaussian model features respectively.  There are a total of
26 free parameters, but 16 of these are associated with the Gaussian
components.  The free parameters are the effective redshift of the
BN11 model ($z_{\rm R}$), the normalization of the intrinsic power-law
continuum, its photon index ($\Gamma$), $N_{H,1}$, $N_{H}$ (radial
column density), $X_{\rm Fe}$ (Fe abundance relative to solar),
$f_{s}$, $A_{\sc apec}$ (normalization of the APEC component),
$kT_{\sc apec}$ (temperature of the APEC component), the FWHM of the
\fekalfa line, the normalizations of the Gaussian emission/absorption
components, and the center energies of the Gaussian
emission/absorption components. The widths of the Gaussian components
could not be constrained by the data and were fixed at $100 \rm \ km
\ s^{-1}$ FWHM, a value much less than the XIS spectral resolution.

The results are shown in \tablespfitsp, from which it can be seen that
the reduced $\chi^{2}$ value for the fit is 1.284, with a null
probability of $4.55 \times 10^{-4}$. Fig.~\ref{fig:szsphuspec} shows
the data and best-fitting unfolded photon
spectrum. Fig.~\ref{fig:szsphdatrat}(a) shows the best-fitting model
overlaid on the counts spectrum, and the corresponding data/model
ratios are shown in Fig.~\ref{fig:szsphdatrat}(c). A zoomed view of
the best-fitting model overlaid on the counts spectrum in the Fe~K
region is shown in Fig.~\ref{fig:szsphdatrat}(b), and the
corresponding data/model ratios are shown in
Fig.~\ref{fig:szsphdatrat}(d). It can be seen that the model provides
an excellent fit to the data. In particular, the simple uniform
spherical model of the absorption, fluorescent emission lines, and
Compton-reflection continuum self-consistently accounts for the
\fekalfap, \fekbetap, the neutral Fe~K edge, and continuum shape, and
this is driven essentially by only two parameters, namely, $N_{H}$,
and $X_{\rm Fe}$. The only other components in the 5--8~keV band shown
in Fig.~\ref{fig:szsphdatrat}(b) and Fig.~\ref{fig:szsphdatrat}(d) are
weak narrow absorption lines due to \fexxvresp, \fexxvip, some
additional \nika line emission, and a weak unidentified feature at
$\sim6.8$~keV (see \tablespfitsp). The best-fitting column density of
the spherical model is $N_{H}=25.78^{+0.19}_{-0.18} \times 10^{22}
\ \rm cm^{-2} $, and $X_{\rm Fe}=1.102^{+0.024}_{-0.021}$.  Both of
these values are unremarkable, the column density indicating a source
that is globally Compton-thin, and the Fe abundance is close to the
Anders and Grevesse (1989) solar value.  The difference is less than
the systematic differences in solar Fe abundance that are reported in
the literature (e.g. see Asplund \etal 2009). What {\it is} remarkable
is that the model versus data plot in Fig.~\ref{fig:szsphdatrat}(b)
and the fit residuals in Fig.~\ref{fig:szsphdatrat}(d) show that the
data appear to leave no room for any broadened relativistic \fekalfa
line emission, and adding such a component to the model would only be
fitting noise. Nevertheless, in \S\ref{relline} we present the results
of adding reflection from a relativistic ionized disk to the model, in
order to obtain upper limits on the flux of such a component.

The best-fitting value of $N_{\rm H,1}$ is $16.73_{-0.65}^{+0.64}
\times 10^{20} \rm \ cm^{-2}$ (\tablespfitsp), which is more than two
orders of magnitude less than the column density of the spherical
model component, $N_{\rm H}$.  The value of $f_{s}$ obtained is
$14.90_{-0.45}^{+0.50} \times 10^{-3}$.  Both $N_{\rm H,1}$ and
$f_{s}$ are small enough that the self-consistency of the absorption,
Compton scattering, and fluorescent line emission from the spherical
model is not broken.  It can be seen from Fig.~\ref{fig:szsphuspec}
that the thermal optically-thin emission and the optically-thin
scattered power-law continuum make a negligible contribution to the
spectrum above $\sim3$~keV, so these components of the model do not
affect the fit parameters of the spherical reprocessor model. The
temperature of the APEC component obtained from the fit is $kT_{\sc
  apec}=0.822_{-0.017}^{+0.016}$~keV.  These values of $kT_{\sc apec}$
and $f_{s}$ are consistent with historical results for extended
emission in \srcp, as well as for similar AGN (see references in the
appendix).

\tablespfits shows that the \fekalfa line flux and EW were measured to
be $9.0 \ \times 10^{-5} \rm \ photons \ cm^{-2} \ s^{-1}$, and
$257$~eV respectively, where the EW was calculated with respect to the
total continuum at the line peak (in the AGN frame).  Statistical
errors on these two quantities were not calculated because the
\fekalfa line in the model is not separable, but the percentage errors
will be similar to the corresponding errors for the \mytorus fits
described in \S\ref{mytorusfits}.  Those errors are of the order of
4\%, on both the line flux and EW (see \tablespfitsp).  The FWHM of
the \fekalfa line from the spherical model fit is $2855^{+725}_{-890}
\ {\rm km \ s^{-1}}$, and this is entirely consistent with the
\chandra HEG measurement of $2430^{+620}_{-590} \ {\rm km \ s^{-1}}$
reported in Shu \etal (2011).  We can estimate the characteristic
radius of the line-emitting region using the simple prescription of
Netzer (1990), and a black-hole mass for \src of $1.58 \times 10^{7}
M_{\odot}$ (Woo and Urry 2002).  This gives $r \sim0.011$~pc and $r
\sim0.015$~pc for the \suzaku and \chandra HEG \fekalfa line FWHM
values respectively. This distance range corresponds to $\sim1.5-2
\times 10^{4}$ gravitational radii.

\begin{table}
\begin{minipage}{.9\textwidth}
\caption[Spectral-Fitting Results for Three X-ray Reprocessor Models]{Spectral-Fitting Results for Three X-ray Reprocessor Models}
\begin{center}
\begin{tabular}{lccc}
\hline
Parameter & Spherical & \mytorus & \mytorus \\
& (BN11) & (Coupled) & (Decoupled) \\
\hline 
$\chi^{2}$ & 409.50 & 398.32 & 396.47 \\
Degrees of Freedom & 319 & 318 & 318 \\
Free Parameters & 26 & 27 & 27 \\
Reduced $\chi^{2}$ & 1.284 & 1.253 &  1.247 \\
Null probability & $4.55\times 10^{-4}$ & $1.46\times 10^{-3}$ & $1.79\times 10^{-3}$ \\
$10^{3} z_{\rm R}$ & $5.32^{+0.43}_{-0.42}$ & $5.80^{+0.37}_{-0.50}$ & $5.30^{+0.42}_{-0.41}$ \\ 
$\Gamma$ & $1.513^{+0.006}_{-0.009}$ & $1.475^{+0.009}_{-0.006}$ & $1.502^{+0.007}_{-0.008}$ \\
$N_{\rm H,1}$ ($10^{20} \ {\rm cm^{-2}}$) & $16.73^{+0.64}_{-0.65}$ & $25.6^{+1.6}_{-1.1}$ & $29.85^{+1.65}_{-0.95}$ \\
$N_{\rm H}$ [sphere] ($10^{22} \ {\rm cm^{-2}}$) & $25.78^{+0.19}_{-0.18}$ & \ldots & \ldots \\
\nhs ($10^{22} \ {\rm cm^{-2}}$) & \ldots & $32.05^{+0.25}_{-0.35}$ & $60.0^{+3.7}_{-3.7}$ \\
\nhz ($10^{22} \ {\rm cm^{-2}}$) & \ldots & $=$ \nhs & $29.24^{+0.29}_{-0.24}$ \\
$A_{\rm S}$ &  \ldots & $2.52^{+0.10}_{-0.08}$ & $1.28^{+0.05}_{-0.04}$ \\
Fe abundance (ratio to solar)  & $1.102^{+0.024}_{-0.021}$ & 1.0(f) & 1.0(f) \\
\thetaobs ($^{\circ}$) & \ldots & $77.20^{+0.40}_{-0.60}$ & \ldots \\
$10^{3}f_{s}$ & $14.90^{+0.50}_{-0.45}$ & $12.71^{+0.45}_{-0.40}$ & $10.15^{+0.46}_{-0.31}$ \\ 
$kT_{\rm apec}$(keV) & $0.822^{+0.016}_{-0.017}$ & $0.810^{+0.020}_{-0.018}$ & $0.802^{+0.018}_{-0.018}$ \\
$A_{\rm apec}$ ($\rm 10^{5} \ ph. \ cm^{-2} \ s^{-1} \ keV^{-1}$) & $12.884^{+0.436}_{-0.434}$ & $15.343^{+0.517}_{-0.613}$  & $18.322^{+0.548}_{-0.732}$ \\ 
\fekalfa $v_{\rm shift}$ ($\rm km \ s^{-1}$) & $-916^{+128}_{-124}$ & $-775^{+110}_{-148}$ & $-923^{+125}_{-122}$ \\ 
\fekalfa FWHM ($\rm km \ s^{-1}$) & $2855^{+725}_{-890}$ &  $3415^{+740}_{-590}$ & $3415^{+460}_{-1295}$ \\
$I_{\rm Fe~K\alpha}$ ($\rm 10^{-5} \ photons \ cm^{-2} \ s^{-1}$) & $8.98$ & $8.66^{+0.34}_{-0.27}$ & $8.66^{+0.34}_{-0.27}$\\
${\rm EW}_{\rm Fe~K\alpha}$ (eV) & $257$ & $255^{+10}_{-8}$ & $255^{+10}_{-8}$ \\
$E_{1}$ (keV) & $0.668^{+0.004}_{-0.005}$ & $0.669^{+0.005}_{-0.006}$ & $0.668^{+0.006}_{-0.005}$ \\
$I_{1}$ ($\rm 10^{-5} \ photons \ cm^{-2} \ s^{-1}$) & $8.97^{+1.05}_{-1.04}$ & $2.79^{+0.51}_{-0.39}$ & $3.00^{+0.50}_{-0.50}$ \\
${\rm EW}_{1}$ (eV) & $504^{+59}_{-58}$ & $157^{+29}_{-22}$ & $182^{+30}_{-30}$ \\ 
$E_{2}$ (keV) & $0.749^{+0.005}_{-0.007}$ & $0.748^{+0.008}_{-0.007}$ & $0.748^{+0.008}_{-0.007}$ \\
$I_{2}$ ($\rm 10^{-5} \ photons \ cm^{-2} \ s^{-1}$) & $3.67^{+0.65}_{-0.65}$ & $1.42^{+0.30}_{-0.31}$ & $1.43^{+0.39}_{-0.24}$ \\
${\rm EW}_{2}$ (eV) & $238^{+42}_{-42}$ & $92^{+19}_{-20}$ & $100^{+28}_{-17}$ \\
$E_{3}$ (keV) & $1.056^{+0.009}_{-0.009}$ & $1.054^{+0.009}_{-0.009}$ & $1.054^{+0.009}_{-0.009}$ \\
$I_{3}$ ($\rm 10^{-5} \ photons \ cm^{-2} \ s^{-1}$) & $1.13^{+0.30}_{-0.19}$ & $0.73^{+0.19}_{-0.15}$ & $0.72^{+0.18}_{-0.15}$ \\
${\rm EW}_{3}$ (eV) & $113^{+30}_{-19}$ & $73^{+19}_{-15}$ & $78^{+19}_{-16}$ \\
$E_{4}$ (keV) & $3.117^{+0.050}_{-0.127}$ & $3.003^{+0.057}_{-0.053}$ & $3.002^{+0.052}_{-0.058}$ \\
$I_{4}$ ($\rm 10^{-5} \ photons \ cm^{-2} \ s^{-1}$) & $0.144^{+0.073}_{-0.076}$ & $0.144^{+0.076}_{-0.074}$ & $0.137^{+0.073}_{-0.077}$ \\
${\rm EW}_{4}$ (eV) & $23.6^{+12.1}_{-12.5}$ & $27.8^{+14.8}_{-14.3}$ & $26.3^{+14.2}_{-14.8}$ \\
$E_{5}$ (keV) & $6.721^{+0.031}_{-0.031}$ & $6.722^{+0.038}_{-0.37}$ & $6.722^{+0.027}_{-0.048}$ \\ 
$I_{5}$ ($\rm 10^{-5} \ photons \ cm^{-2} \ s^{-1}$) & $-0.544^{+0.174}_{-0.172}$ & $-0.48^{+0.16}_{-0.19}$ & $-0.51^{+0.17}_{-0.18}$ \\
${\rm EW}_{5}$ (eV) & $-15.6^{+5.0}_{-4.9}$ & $-13.6^{+4.5}_{-5.5}$ & $-14.4^{+4.8}_{-5.2}$ \\
$E_{6}$ (keV) & $6.822^{+0.032}_{-0.037}$ & $6.832^{+0.029}_{-0.033}$ & $6.832^{+0.033}_{-0.033}$ \\
$I_{6}$ ($\rm 10^{-5} \ photons \ cm^{-2} \ s^{-1}$) & $0.531^{+0.180}_{-0.187}$ & $0.62^{+0.15}_{-0.22}$ & $0.56^{+0.17}_{-0.20}$ \\
${\rm EW}_{6}$ (eV) & $15.1^{+5.1}_{-5.3}$ & $17.6^{+4.2}_{-6.3}$ & $15.7^{+5.7}_{-4.7}$ \\
$E_{7}$ (keV) & $7.020^{+0073}_{-0.077}$ & $6.991^{+0.070}_{-0.109}$ & $6.970^{+0.078}_{-0.077}$ \\
$I_{7}$ ($\rm 10^{-5} \ photons \ cm^{-2} \ s^{-1}$) & $-0.289^{+0.187}_{-0.189}$ & $-0.24^{+0.19}_{-0.19}$ & $-0.26^{+0.17}_{0.20-}$ \\
${\rm EW}_{7}$ (eV) & $7.0^{+4.0}_{-6.1}$ & $-5.4^{+4.3}_{-4.2}$ & $-6.5^{+5.0}_{-4.3}$ \\
$E_{8}$ (keV) & $7.487^{+0.049}_{-0.041}$ & $7.489^{+0.030}_{-0.029}$ & $7.489^{+0.028}_{-0.026}$ \\
$I_{8}$ ($\rm 10^{-5} \ photons \ cm^{-2} \ s^{-1}$) & $0.372^{+0.178}_{-0.180}$ & $0.65^{+0.20}_{-0.17}$ & $0.70^{+0.16}_{-0.20}$ \\
${\rm EW}_{8}$ (eV) & $16.4^{+7.9}_{-7.9}$ & $28.9^{+8.7}_{-7.7}$ & $31.2^{+9.4}_{-7.1}$ \\
\hline
\end{tabular}
\end{center} 
Spectral-fitting results for the \suzaku data for \srcp, with a uniform spherical model of the X-ray reprocessor (BN11),
a toroidal model with \mytorusp, fitted in coupled mode, and
a model with \mytorus fitted in decoupled mode. See text for details. Fixed parameters are indicated by (f). The parameter $z_{\rm R}$
for X-ray reprocessing model components is fitted independently of the cosomological redshift, and it is
used to derive the apparent velocity shift of the \fekalfa line peak, $v_{\rm shift}$. Parameter values for all other components
are rest-frame values. The photon flux and EW of the \fekalfa line are denoted by the parameters $I_{\rm Fe~K\alpha}$ and ${\rm EW}_{\rm Fe~K\alpha}$ respectively. Note that for the BN11 model, the \fekalfa line flux cannot be derived independently of the continuum, so the \fekalfa line flux and EW are estimates only, and do not have statistical errors. The parameters $I_{n}$, and ${\rm EW}_{n}$, where $n=1$ to $8$, denote the photon fluxes and EW values respectively, for the additional Gaussian emission or absorption lines that are included in all of the models. Parameters that have no entry in a particular column are not part of the model in that column. 
\end{minipage}
\end{table} 

\begin{table}
\begin{minipage}{.9\textwidth}
\caption[Fluxes and Luminosities from X-ray Reprocessor Spectral Fits]{Fluxes and Luminosities from X-ray Reprocessor Spectral Fits}

\begin{center}
\begin{tabular}{lccc}
\hline
 & Spherical & \mytorus & \mytorus \\
& (BN11) & (Coupled) & (Decoupled) \\
\hline 
$F_{\rm obs}$[0.5--2 keV] ($10^{-12} \rm \ erg \ cm^{-2} \ s^{-1}$) & 0.475 & 0.488 & 0.489 \\
$F_{\rm obs}$[2--10  keV] ($10^{-12} \rm \ erg \ cm^{-2} \ s^{-1}$) & 19.5 & 19.5 & 19.5 \\
$F_{\rm obs}$[10--20  keV] ($10^{-12} \rm \ erg \ cm^{-2} \ s^{-1}$) & 36.0 & 36.2 & 36.2 \\
$L_{\rm obs}$[0.5--2 keV] ($10^{40} \rm \ erg \ s^{-1}$) & 6.73 & 7.44 & 7.50 \\
$L_{\rm obs}$[2--10 keV] ($10^{40} \rm \ erg \ s^{-1}$) & 274.9 & 274.6 & 274.4 \\
$L_{\rm obs}$[10--20 keV] ($10^{40} \rm \ erg \ s^{-1}$) & 510.4 & 513.4 & 513.6 \\
$L_{\rm intr}$[0.5--2 keV] ($10^{40} \rm \ erg \ s^{-1}$) & 315.7 & 292.1 & 334.4 \\
$L_{\rm intr}$[2--10 keV] ($10^{40} \rm \ erg \ s^{-1}$) & 752.9 & 736.9 & 806.0 \\
$L_{\rm intr}$[10--20 keV] ($10^{40} \rm \ erg \ s^{-1}$) & 556.3 & 570.0 & 602.4  \\
\hline
\end{tabular}
\end{center} 
Fluxes and luminosities corresponding to the spectral-fitting results shown in \tablespfitsp, for the three X-ray
reprocessor models, as described in the text.
$F_{\rm obs}$ and $L_{\rm obs}$ are observed fluxes and luminosities, respectively. Intrinsic continuum luminosities,
inferred by removing all absorption and scattering components, are denoted by $L_{\rm intr}$. 
\end{minipage}
\end{table} 

\begin{table}
\begin{minipage}{.9\textwidth}
\caption[Relativistic Disk Model Fractional Contribution Upper Limits]{Relativistic Disk Model Fractional Contribution Upper Limits}
\begin{center}
\begin{tabular}{lllcccc}
\hline

$\cos{\theta_{\rm obs}}$ & $R_{X}^{a}$ & $\log{\xi}$ & $f_{\rm UR}$ (a=0.0) & $f_{\rm UR}$ (a=0.9982) & $f_{\rm UR}$ (a=0.0) & $f_{\rm UR}$ (a=0.9982) \\
& & & broadband & broadband & ($>3$ keV) & ($>3$ keV)\\

0.95 & 0.5  &   0.0    & 0.95026 & 1.01198 & 1.13870 & 1.36835 \\
0.95 & 0.5  &   1.0    & 0.95486 & 1.02123 & 1.14640 & 1.38676 \\
0.95 & 0.5  &   2.0    & 0.98103 & 1.05204 & 1.18192 & 1.45730 \\
0.95 & 0.5  &   3.0    & 0.90902 & 0.93055 & 1.02292 & 1.19616 \\
0.95 & 0.5  &   4.0    & 0.79524 & 0.79525 & 0.70773 & 0.68441 \\
0.95 & 1.0  &   0.0    & 0.96188 & 1.07286 & 1.50038 & 2.09108 \\
0.95 & 1.0  &   1.0    & 0.96955 & 1.08824 & 1.49876 & 2.07085 \\
0.95 & 1.0  &   2.0    & 1.00952 & 1.13436 & 1.39816 & 1.90382 \\
0.95 & 1.0  &   3.0    & 0.91801 & 0.96112 & 0.99252 & 1.13419 \\
0.95 & 1.0  &   4.0    & 0.78598 & 0.78756 & 0.75590 & 0.74346 \\
0.95 & 2.0  &   0.0    & 0.93796 & 1.10267 & 1.19436 & 1.61013 \\
0.95 & 2.0  &   1.0    & 0.95179 & 1.12518 & 1.20829 & 1.64392 \\
0.95 & 2.0  &   2.0    & 0.99630 & 1.19572 & 1.26090 & 1.76053 \\
0.95 & 2.0  &   3.0    & 0.91929 & 0.98395 & 1.00714 & 1.20027 \\
0.95 & 2.0  &   4.0    & 0.78602 & 0.77517 & 0.70306 & 0.68443 \\
0.95 & 3.0  &   0.0    & 0.91601 & 1.10831 & 1.51258 & 2.06379 \\
0.95 & 3.0  &   1.0    & 0.92520 & 1.12518 & 1.50479 & 2.04676 \\
0.95 & 3.0  &   2.0    & 0.97882 & 1.19572 & 1.40267 & 1.89232 \\
0.95 & 3.0  &   3.0    & 0.91143 & 0.98996 & 1.01585 & 1.16683 \\
0.95 & 3.0  &   4.0    & 0.78603 & 0.77675 & 0.75435 & 0.73879 \\
0.05 & 0.5  &   0.0    & 1.08170 &  1.10789 & 1.29876 & 1.36906 \\
0.05 & 0.5  &   1.0    & 1.08322 &  1.10632 & 1.31105 & 1.38280 \\
0.05 & 0.5  &   2.0    & 1.04314 &  1.06319 & 1.36644 & 1.46093 \\
0.05 & 0.5  &   3.0    & 0.92045 &  0.95921 & 1.11891 & 1.22387 \\
0.05 & 0.5  &   4.0    & 0.82301 &  0.81990 & 0.69373 & 0.68910 \\
0.05 & 1.0  &   0.0    & 1.15345 &  1.23235 & 1.84501 & 2.10342 \\
0.05 & 1.0  &   1.0    & 1.15494 &  1.22924 & 1.88070 & 2.09555 \\
0.05 & 1.0  &   2.0    & 1.10113 &  1.16186 & 1.71892 & 1.93663 \\
0.05 & 1.0  &   3.0    & 0.94212 &  1.03043 & 1.06880 & 1.15754 \\
0.05 & 1.0  &   4.0    & 0.82148 &  0.82616 & 0.75125 & 0.75128 \\
0.05 & 2.0  &   0.0    & 1.18648 &  1.23235 & 1.42997 & 1.65975 \\
0.05 & 2.0  &   1.0    & 1.18797 &  1.22924 & 1.45310 & 1.69654 \\
0.05 & 2.0  &   2.0    & 1.12354 &  1.16186 & 1.54558 & 1.82209 \\
0.05 & 2.0  &   3.0    & 0.96379 &  1.03043 & 1.10767 & 1.23877 \\
0.05 & 2.0  &   4.0    & 0.82461 &  0.82616 & 0.68750 & 0.96706 \\
0.05 & 3.0  &   0.0    & 1.18604 &  1.23182 & 1.89572 & 2.07447 \\
0.05 & 3.0  &   1.0    & 1.18598 &  1.22564 & 1.88018 & 2.05900 \\
0.05 & 3.0  &   2.0    & 1.12010 &  1.16295 & 1.72010 & 1.92037 \\
0.05 & 3.0  &   3.0    & 0.97152 &  1.04745 & 1.09835 & 1.21041 \\
0.05 & 3.0  &   4.0    & 0.82306 &  0.82772 & 0.74657 & 0.75128 \\
\hline
\end{tabular}
\end{center} 
The upper limits on the percentage contribution of relativistic disk
reflection to the \suzaku spectra of \srcp, when the disk reflection
is modelled with \relxillp, and included with a uniform spherical
model for the distant-matter Compton reflection and Fe~K line emission
(see text for details). Shown are values of $f_{\rm UR}$, defined as
the upper limit on the percentage of the 2--20~keV total observed flux
that is due to only the \relxill component, for a given set of
\relxill model parameters. The values of $f_{\rm UR}$ are shown for 80
combinations of the key \relxill parameters, $a$ (dimensionless
black-hole spin), cosine of the inclination angle ($\cos{\theta_{\rm
    obs}}$), reflection fraction ($R_{X}$), and the logarithm of the
ionization parameter (\logxip). The corresponding values of the
remaining \relxill model parameters are given in the text. Columns 4
and 5 show values of $f_{\rm UR}$ obtained from the broadband spectral
fits, whereas columns 6 and 7 show the values of $f_{\rm UR}$ obtained
from the fits performed on data above $3$~keV. In the latter case, the
values of $f_{\rm UR}$ were calculated by extrapolating each model
down to 2 keV.
\end{minipage}
\end{table}

\tablespfits shows that the photon index of the intrinsic power-law
continuum for the spherical model fit is $\Gamma =
1.513^{+0.006}_{-0.009}$, which is rather flat with respect to
characteristic values of $\Gamma \sim1.8$--$2.0$. However, statistical
studies of AGN (both type~1 and type~2) do show a a range in $\Gamma$
of 1.5--2.5 (e.g., Dadina 2007, 2008; Balokovi\'{c} \etal 2020, and
references therein).  In addition, these studies show a non-negligible
number of outliers outside of this range.

Continuum fluxes and luminosities obtained from the spherical model
fit are shown in \tablefluxesp, in three energy bands: 0.5--2~keV,
2--10~keV, and 10--20~keV.  These energy bands refer to the observed
frame for fluxes, and to the source frame for luminosities. The
0.5--2~keV, 2--10~keV, and 10--20~keV fluxes are $0.475$, $19.5$, and
$36.0 \ \times \ 10^{-12} \rm \ erg \ cm^{-2} \ s^{-1}$ respectively.
Observed and intrinsic luminosities are denoted by $L_{\rm obs}$ and
$L_{\rm intr}$ respectively.  The 0.5--2~keV, 2--10~keV, and
10--20~keV values for $L_{\rm obs}$ that we obtained are $0.067$,
$2.75$, and $5.10 \ \times 10^{42} \ \rm erg \ s^{-1}$
respectively. The intrinsic luminosities in the 0.5--2~keV, 2--10~keV,
and 10--20~keV bands are $3.16$, $7.53$, and $5.56 \ \times 10^{42}
\ \rm erg \ s^{-1}$ respectively. Comparison of the intrinsic
luminosities with corresponding values in the literature is
complicated by the fact that they are often derived from \adhoc
models, which can easily yield very different luminosities.

\subsubsection{The Additional Gaussian Components}
\label{szsphsoftlines}

\begin{figure}
\centerline{
        \includegraphics[width=6cm,angle=270]{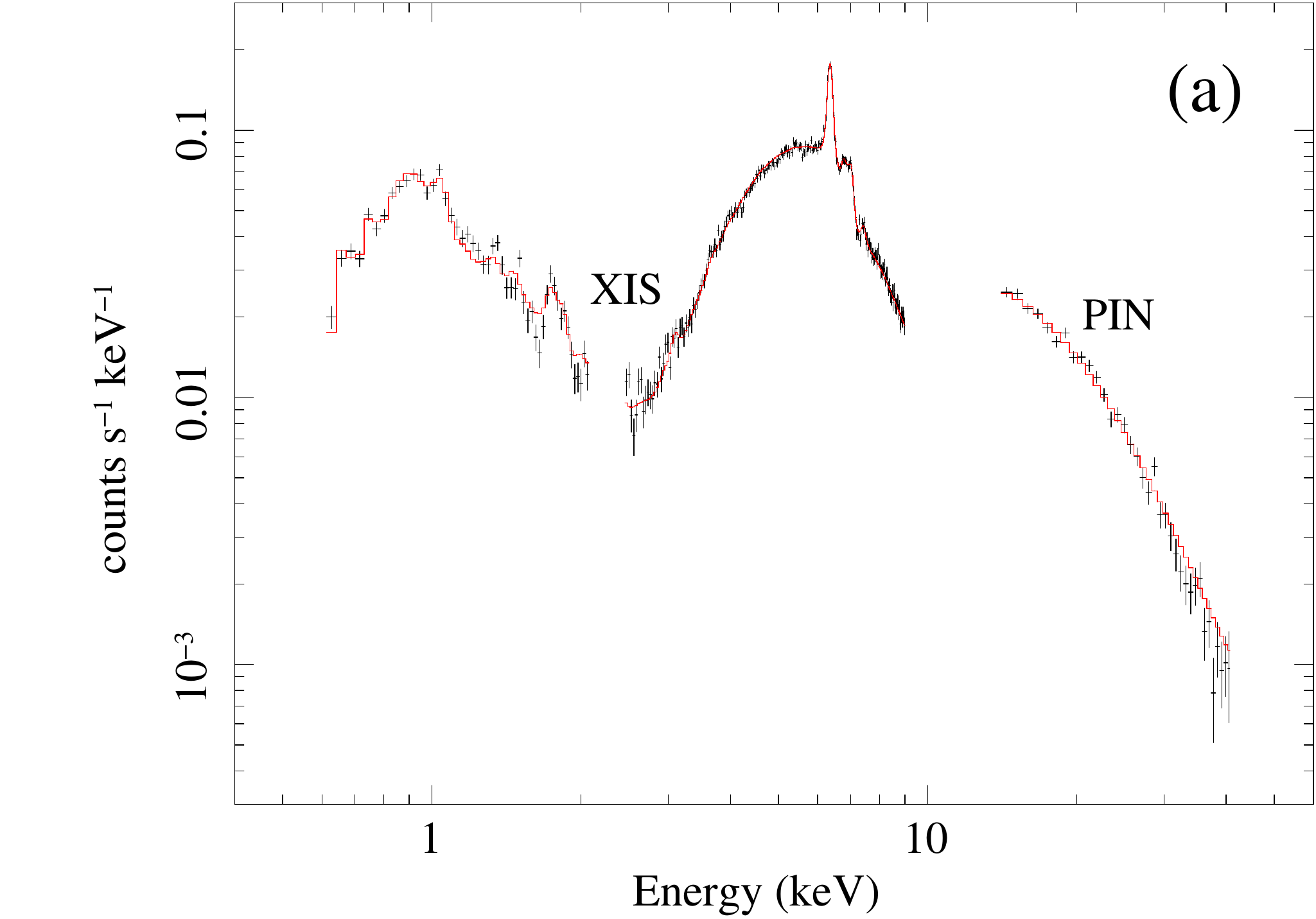}
        \includegraphics[width=6cm,angle=270]{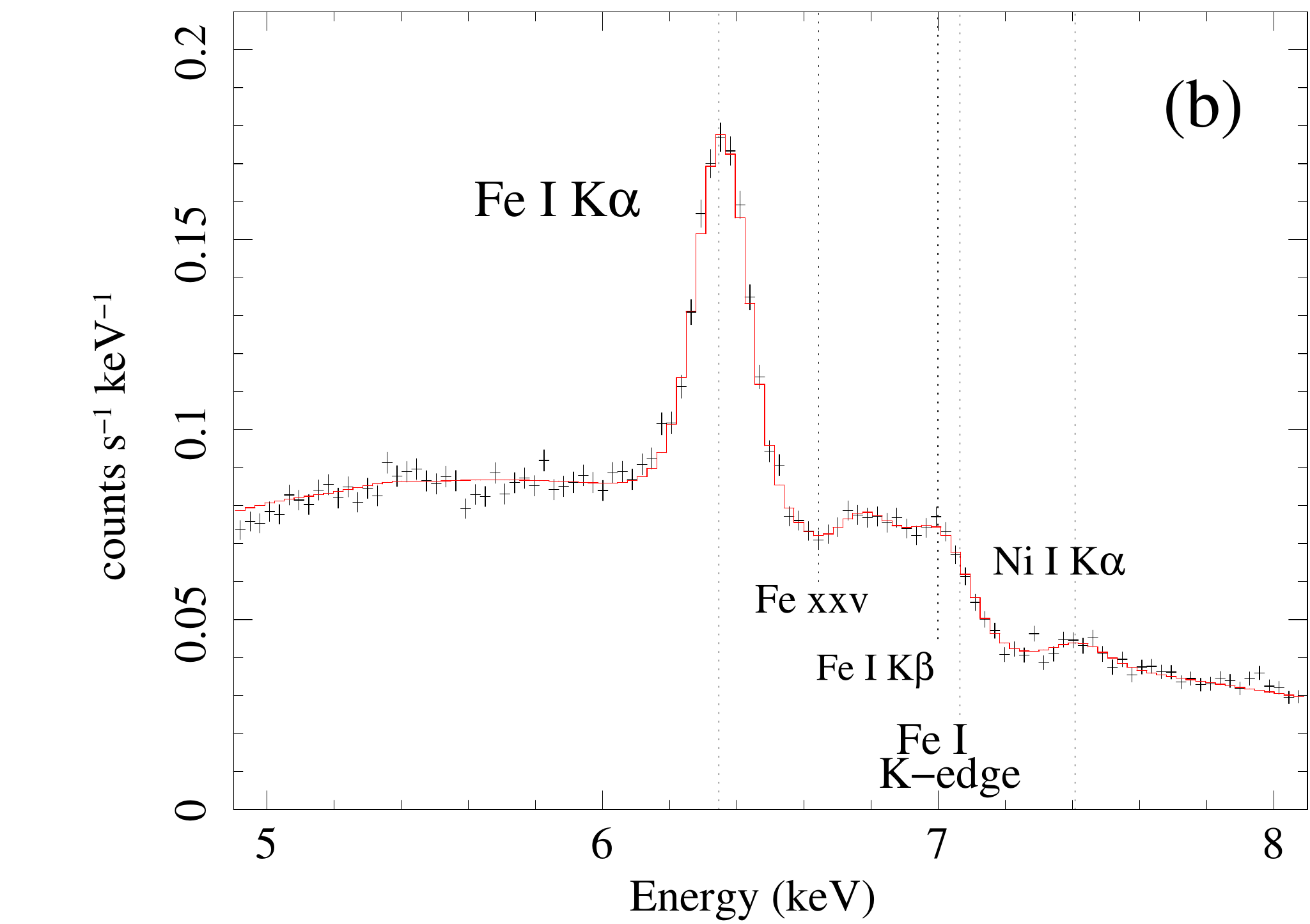}\hspace{1mm}
        }
\centerline{
        \includegraphics[width=3.43cm,angle=270]{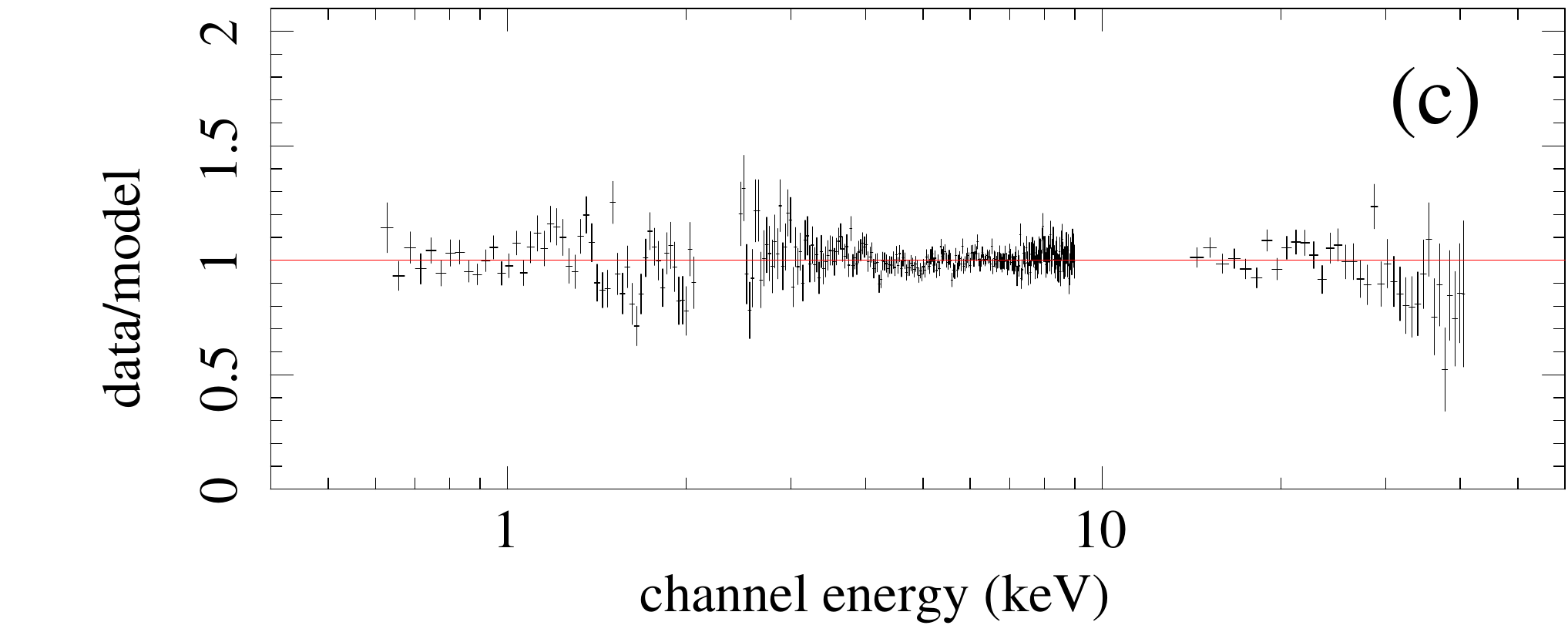}
        \includegraphics[width=3.43cm,angle=270]{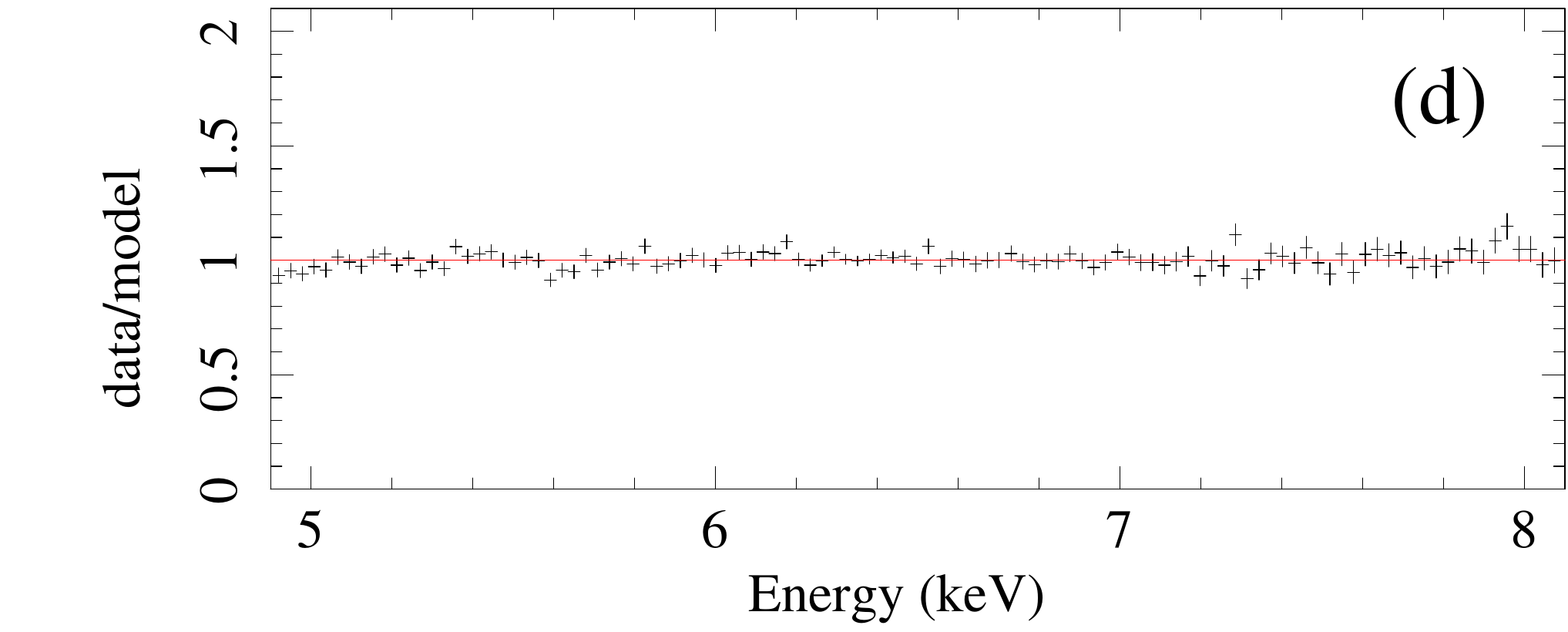}
        } 
        \caption{\small Same broadband spectral fit as in
          Fig.~\ref{fig:szsphuspec} (see \S\ref{sphericalfit} and
          \tablespfitsp), showing the uniform spherical X-ray
          reprocessor model fitted to the \src \suzaku spectra.  (a)
          The full broadband XIS and PIN data (black) overlaid with
          the best-fitting spherical (BN11) model (red).  Caption text
          for panels (b), (c), and (d) as in Yaqoob \etal (2015):
          ``(b) The data and model shown in (a), zoomed in on the Fe~K
          region, showing the detailed fit to the \fekalfa emission
          line and other atomic features, as labelled. Note that the
          flux of the model \fekalfa line is not controlled by an
          arbitrary parameter, but calculated self-consistently from
          the same matter distribution that produces the global
          Compton-scattered continuum. The dotted lines correspond to
          the expected energies of the labelled atomic features in the
          observed frame. The data to model ratios corresponding to
          (a) and (b) are shown in (c) and (d) respectively''. Note
          that zoomed plots are shown specifically for the Fe~K band,
          as opposed to the soft X-ray spectrum, because the purpose
          of the present paper is to constrain the matter distribution
          that produces the Fe~K fluorescent lines, the line-of-sight
          extinction, and the Compton-scattered continuum. The
          spectral signatures from that material are insignificant
          below $\sim2$~keV, (see Fig.~\ref{fig:szsphuspec}), and the
          soft X-ray spectrum originates in different material.
          Consequently, the constraints on the high-energy spectrum
          are not sensitive to the details of how the soft X-ray
          spectrum is modelled.  }
\label{fig:szsphdatrat}
\end{figure}

\tablespfits shows the results for the eight Gaussian model components
that were included in the spherical reprocessor model fit. Shown are
the line centroid energies, fluxes, and equivalent widths (EW). There
are six emission lines and two absorption lines (the latter have
negative fluxes in \tablespfitsp). From comparison of the fitted
energies with expected energies, we identify the emission-line
components 1, 3, and 4 with \olyap, \nelyap, and \arlya
respectively. Component 2 is unidentified, as is component 6. We have
already mentioned the Gaussian components 5 to 8, which lie in the
$\sim6$--8~keV energy band, shown in Fig.~\ref{fig:szsphdatrat}(b). We
identify components 5 and 7 with absorption due to \fexxvres and
\feklya respectively. It is possible that all five components
identified with highly ionized atomic species (components 1, 3, 4, 5,
and 7) originate in the same ionized plasma, which may be nonuniform
in ionization state and other physical properties.  The emission lines
would be produced by material out of the line-of-sight, and the two Fe
absorption lines would be produced by material in the line-of-sight,
possibly in the form of a wind. The statistical errors on the centroid
energies are too large to allow a reliable determination of the
outflow velocity of such a wind, but it could be as low as $\sim100
\ \rm km \ s^{-1}$, or as large as $\sim6300 \ \rm km
\ s^{-1}$. Highly ionized species of Fe in outflow have been observed
in other AGN (e.g., Tombesi 2015; Braito \etal 2018, and references
therein).  We do not investigate the ionized outflow further with
physical models, as it is beyond the scope of the present work.
Future observations with the calorimeter aboard \xrism will provide
higher spectral resolution data, which will be better for constraining
the physics and geometry of the outflow.

We identify Gaussian component 8 with \nika fluorescence emission,
which is likely produced in the same neutral matter that produces the
\fekalfa line. Even though the BN11 spherical model already includes
\nika fluorescence emission, the non-zero flux of Gaussian component 8
implies that the BN11 model has a deficit in the \nika line
flux. There is in fact considerable uncertainty in the solar value of
the Ni abundance, so the need for additional \nika line flux in the
model is not unexpected (e.g., see Yaqoob and Murphy 2011).

\subsection{Spectral Fits with the \mytorus Model}
\label{mytorusfits}

The ``coupled'' and ``decoupled'' modes of the \mytorus
spectral-fitting model for X-ray reprocessing (Murphy \& Yaqoob 2009)
have been described extensively in Yaqoob 2012, and LaMassa \etal
2014. The present application for \src follows closely that for Mkn~3
(Yaqoob \etal 2015). Full details of the parameters, assumptions, and
caveats can be found in that study, and we use the same particular
\mytorus model tables. The key parameters of the \mytorus model are
the normalization of the incident power-law continuum ($A_{\rm PL}$),
its photon index ($\Gamma$), the direct line-of-sight column density
(\nhzp), the global column density (\nhsp), and $A_{S}$, the relative
normalization between the direct continuum and the reflected continuum
(and fluorescent line emission).  In coupled mode, by definition, both
column densities are equal to the equatorial column density (e.g.,
Yaqoob \etal 2015), and the parameter \thetaobs denotes the
inclination angle of the line-of-sight to the observer with respect to
the torus symmetry axes.  In decoupled mode, the zeroth-order (i.e.,
direct) continuum is decoupled from the inclination angle (which
becomes a dummy parameter), and the model can then crudely mimic
different non-toroidal geometries. Velocity broadening of the \mytorus
emission components is implemented in the same way as it was for the
BN11 spherical model, and we also include the same additional spectral
components corresponding to optically-thin scattering, optically-thin
thermal emission, and the eight empirical Gaussian components (see
\S\ref{sphericalfit}).  We present the results of fitting the \suzaku
\src spectra with the coupled and decoupled \mytorus models in
\S\ref{coupledmytfits} and \S\ref{decoupmytfits} respectively.

\subsubsection{Coupled \mytorus Model}
\label{coupledmytfits}

The XSPEC model expression for the coupled \mytorus model that was applied to the \src spectra is:

\begin{eqnarray}
\rm
\mytorus \ \ model \  =a \  constant<1>*phabs<2>( & & \nonumber \\ 
\rm zpowerlw<3>*etable\{mytorus\_Ezero\_v00.fits\}<4> & + & \nonumber \\
\rm constant<5>*(gsmoothb<6>(atable\{mytorus\_scatteredH200\_v00.fits\}<7>)) & + & \nonumber \\
\rm constant<8>*(gsmooth<9>(atable\{mytl\_V000010nEp000H200\_v00.fits\}<10>)) & + & \nonumber \\
\rm constant<11>*zpowerlw<12> + \rm zphabs<13>*apec<14> & + & \nonumber \\
\rm \sum_{n=15}^{n=23} { zgauss<n> } ) \\
\end{eqnarray}

\begin{figure}
\centerline{
        \includegraphics[width=6cm,angle=270]{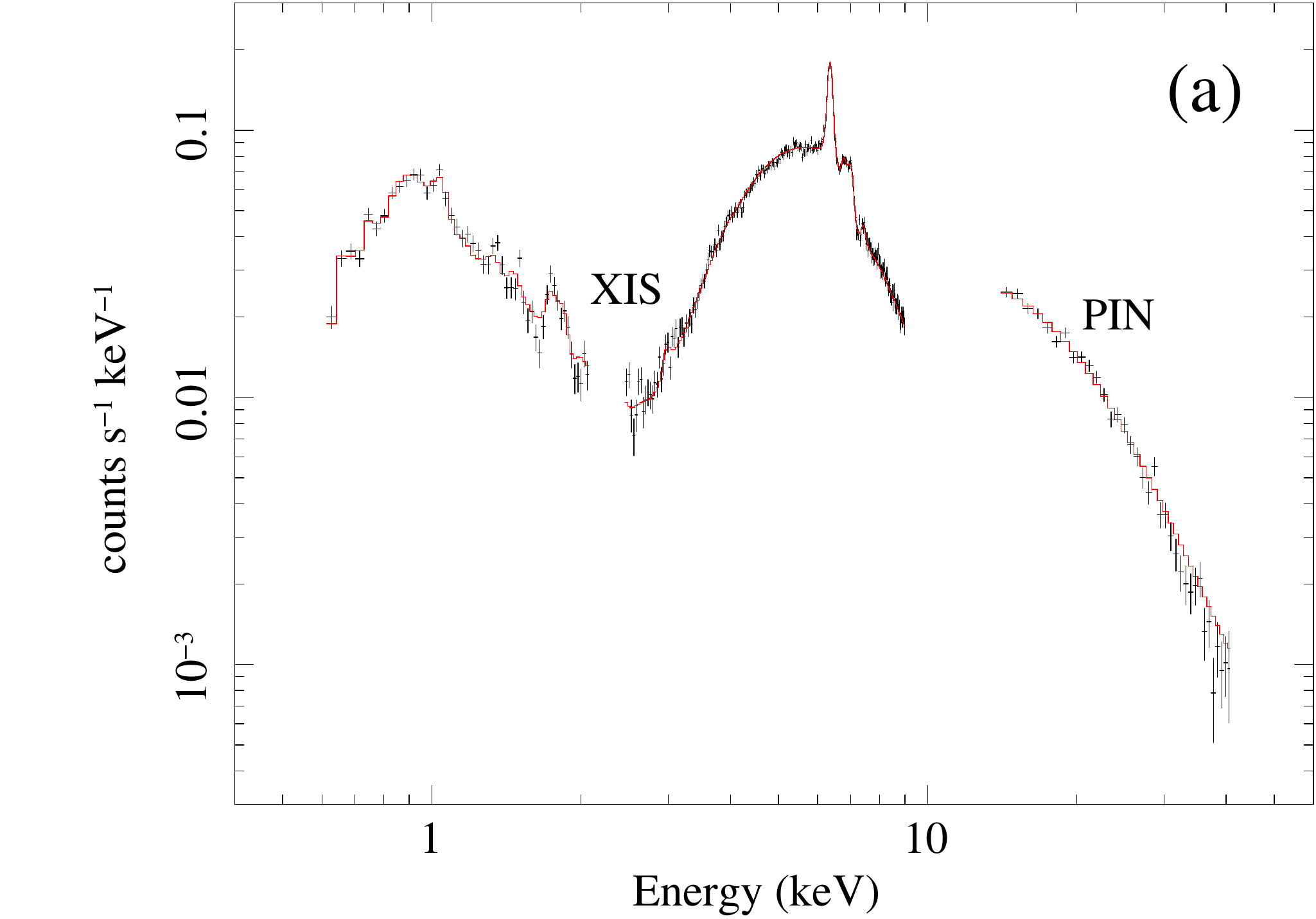}
        \includegraphics[width=6cm,angle=270]{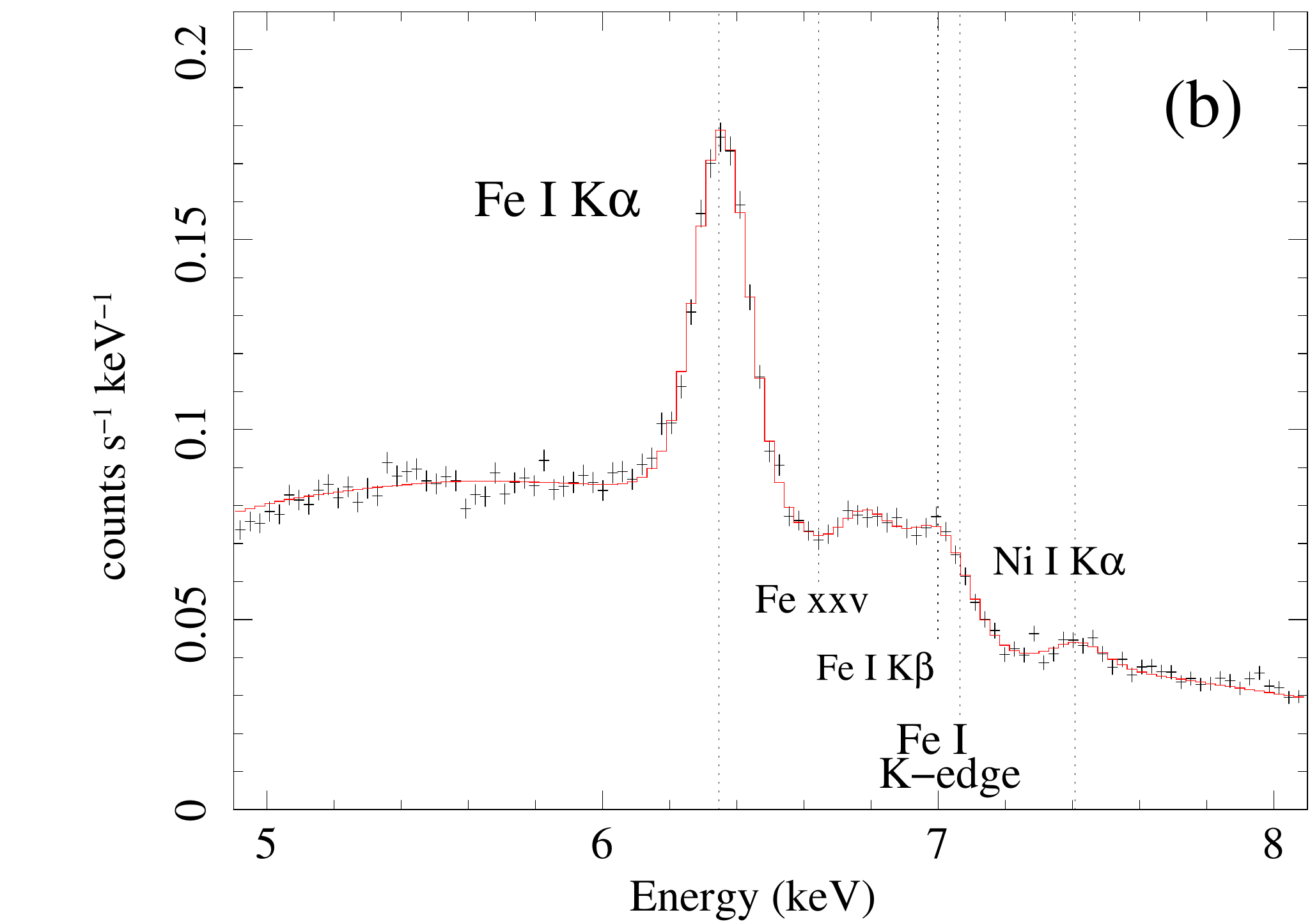}\hspace{1mm}
        }
\centerline{
        \includegraphics[width=3.43cm,angle=270]{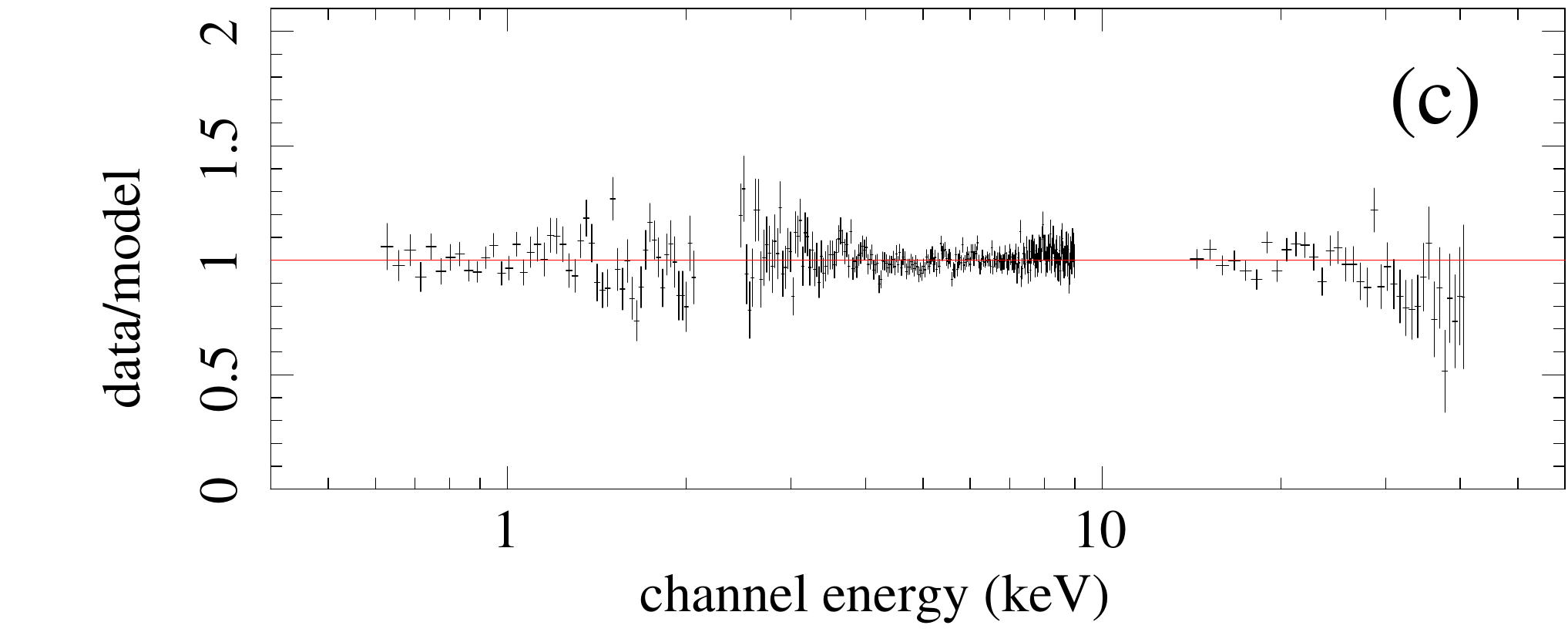}
        \includegraphics[width=3.43cm,angle=270]{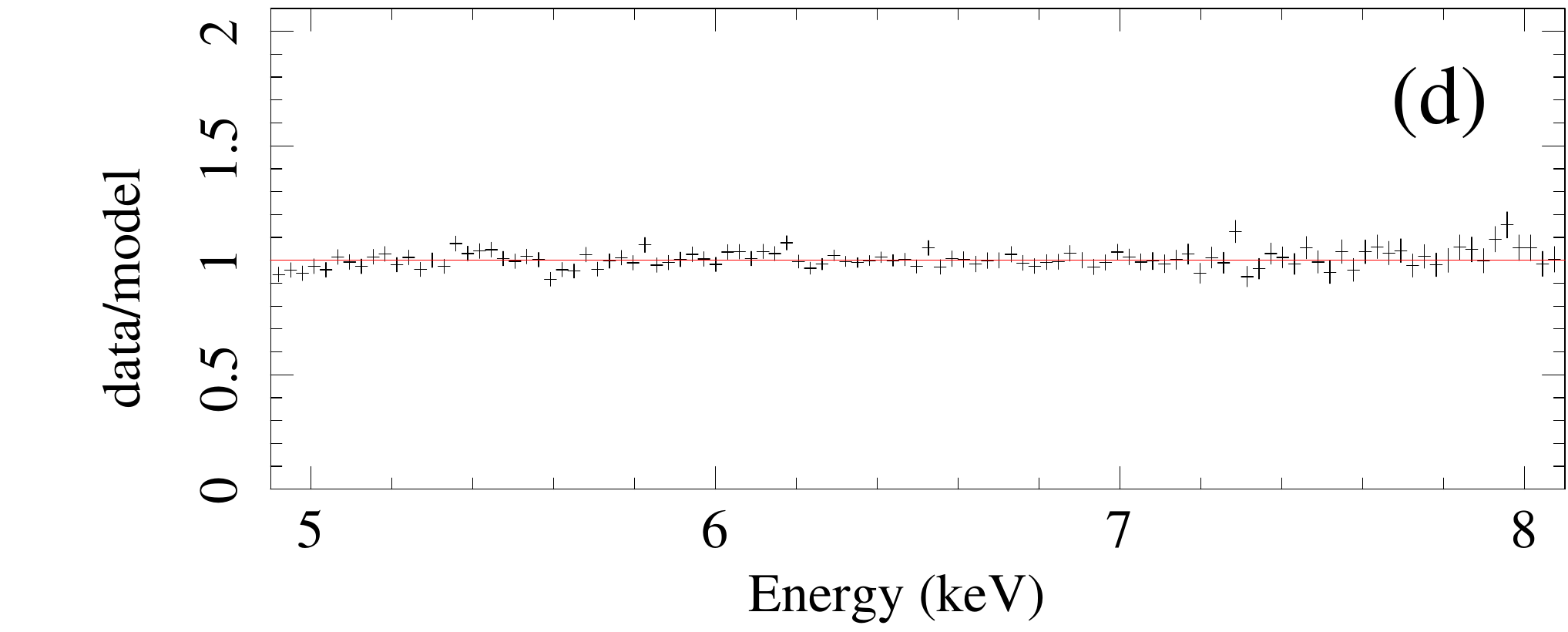}
        }
 \caption{\small  Broadband spectral fit to the \src \suzaku spectra with the
coupled \mytorus model (see \S\ref{coupledmytfits} and \tablespfitsp). All of the descriptions
in the caption for Fig.~\ref{fig:szsphdatrat} apply here.
}
\label{fig:szmytcoupdatrat}
\end{figure}

Here we identify $\rm constant<1> = C_{\rm PIN:XIS}$, $\rm phabs<2> =
$ Galactic column density, ${\rm constant<5>} = A_{S}$ = ${\rm
  constant<8>}$, and ${\rm constant<11>} = f_{s}$. In coupled mode,
the column densities associated with each of the three \mytorus model
tables (components 4, 7, and 10 above) are tied together, as are the
inclination angles.  The corresponding parameters of the two Gaussian
velocity broadening components, ${\rm gsmooth<6>}$ and ${\rm
  gsmooth<9>}$, are tied together. Note that, whereas the BN11
spectral fit included an additional (thin) absorbing screen that
covered all emission components, we found that it was only required
for the optically-thin thermal emission component in the \mytorus
fit. The column density of this component is again denoted by $N_{\rm
  H,1}$ (component 13 on the model expression).  There are a total 27
free parameters in the fit, one more than in the BN11 fit (again, 16
of these parameters are associated with the Gaussian components).  The
spectral-fitting results are given in \tablespfitsp, from which it can
be seen that the $\chi^{2}$ value for the fit is 398.32 (corresponding
to a reduced $\chi^{2}$ value of 1.253), with a null probability of
$1.46 \times 10^{-3}$. Fig.~\ref{fig:szmytcoupdatrat}(a) shows the
best-fitting model overlaid on the counts spectrum, and the
corresponding data/model ratios are shown in
Fig.~\ref{fig:szmytcoupdatrat}(c). A zoomed view of the best-fitting
model overlaid on the counts spectrum in the Fe~K region is shown in
Fig.~\ref{fig:szmytcoupdatrat}(b), and the corresponding data/model
ratios are shown in Fig.~\ref{fig:szmytcoupdatrat}(d). It can be seen
that the model provides an excellent fit to the data, with no
significant residuals in the Fe~K band. Although the \mytorus fit is
formally statistically marginally better than the spherical model fit,
we do not interpret the results as favoring one geometry over the
other. This is because the complex soft X-ray spectral modelling for
both cases needs improvement, in terms of better data and models, and
the most important aspect for the present study is that both models
leave no statistically significant residuals in the critical Fe~K
band.  The inclination angle and column density from the coupled
\mytorus fit are $(77.20^{+0.40}_{-0.60})^{\circ}$ and
$3.205^{+0.025}_{-0.035} \times 10^{23} \ \rm cm^{-2}$
respectively. As noted in Murphy and Yaqoob (2009), this column
density should be multiplied by $\pi/4$ in order to compare with the
radial column density from the BN11 spherical model.  Taking this into
account, the coupled torus model column density is in good agreement
with the column density obtained from the uniform spherical model.
The value of the constant $A_{S}=2.52^{+0.10}_{-0.08}$ is more than
double the value for a time-steady intrinsic X-ray continuum
illuminating a torus with a covering factor of 0.5.  The value of
$f_{s}$ for the coupled \mytorus fit is $12.71^{+0.45}_{-0.40} \times
10^{-3}$, or $\sim15\%$ lower than the corresponding value for the
BN11 fit. The intrinsic power-law continuum is a little flatter than
it is in the BN11 fit, with $\Gamma=1.475^{+0.009}_{-0.006}$.

The magnitude of the shift of the peak of \fekalfa emission line is
$\sim15\%$ lower than it is for the BN11 model, with non-overlapping
statistical errors. However, the flux and EW of the line are similar
for the two models, and statistically consistent with each other. The
FWHM of the \fekalfa line is formally $\sim20\%$ higher from the
coupled \mytorus fit, but it is still statistically consistent with
both the BN11 fit to the \suzaku spectra and the historical \chandra
HETG measurement described in \S\ref{sphericalfit}. In addition, the
centroid energies, fluxes, and EW values of all eight Gaussian line
components are all statistically consistent with the corresponding
values from the BN11 fit, and the same is true of the temperature of
the optically-thin thermal continuum ($kT_{\sc apec}$). However, the
relative normalization of the latter emission component from the
\mytorus fit is $\sim19\%$ higher than it is for the BN11 fit. On the
other hand, all of the net observed fluxes and luminosities agree
within $11\%$ for the BN11 and coupled \mytorus models. The intrinsic
luminosities are not expected to be similar, since they are
model-dependent. Nevertheless, our fits show that they are similar
within $\sim8\%$ for the two models. The column density, $N_{\rm
  H,1}$, associated with the \apec component is $25.6^{+1.6}_{-1.1}
\times 10^{20} \ \rm cm^{-2}$, the same order of magnitude as the
uniform screen applied in the BN11 model.

\subsubsection{Decoupled \mytorus Model}
\label{decoupmytfits}

In decoupled mode, the XSPEC model expression is the same as that for
the coupled mode (\S\ref{coupledmytfits}), except that the column
density for the zeroth-order continuum, $N_{\rm H,Z}$ (associated with
component 4), is independent of the global column density, $N_{\rm
  H,S}$ (associated with components 7 and 10), which is responsible
for producing the Compton-scattered continuum and the fluorescent line
emission respectively.  As explained in Yaqoob (2012), and Yaqoob
\etal (2015), in decoupled mode, the inclination angle associated with
the zeroth-order continuum is a dummy parameter, fixed at
$90^{\circ}$, and for the reflected spectrum the data are not
sensitive to the associated inclination angle if all of the column
densities involved are Compton-thin.  The spectral fits to \src so far
show that all of the column densities are indeed Compton-thin, so we
simply choose to fix the inclination angle associated with the
reflection spectrum and fluorescent line emission tables at
$0^{\circ}$.

The spectral-fitting results for the decoupled model fit are given in
the fourth column of \tablespfitsp, from which it can be seen that the
$\chi^{2}$ value for the fit is 396.47 (corresponding to a reduced
$\chi^{2}$ value of 1.247), with a null probability of $1.79 \times
10^{-3}$. The difference in $\chi^{2}$ between the coupled and
decoupled \mytorus models is $<2$, for the same number of free
parameters, so the decoupled \mytorus fit is essentially statistically
indistinguishable from the coupled fit.
Fig.~\ref{fig:szmytcoupdatrat}(a) shows the best-fitting decoupled
model overlaid on the counts spectrum, and the corresponding
data/model ratios are shown in Fig.~\ref{fig:szmytcoupdatrat}(c). A
zoomed view of the best-fitting model overlaid on the counts spectrum
in the Fe~K region is shown in Fig.~\ref{fig:szmytcoupdatrat}(b), and
the corresponding data/model ratios are shown in
Fig.~\ref{fig:szmytcoupdatrat}(d). It can be seen that the model again
provides an excellent fit to the data, with no significant residuals
in the Fe~K band.  The global column density is $N_{\rm H,S} =
60.0^{+3.7}_{-3.7} \times 10^{22} \ \rm cm^{-2}$, which is a factor of
$\sim2$ larger than the equatorial column density from the coupled
\mytorus fit.  The parameter $A_{S}$ is $1.28^{+0.05}_{-0.04}$, a
factor of $\sim2$ smaller than that from the coupled \mytorus fit. The
column density in the line-of-sight is $N_{\rm H,Z} =
29.24^{+0.29}_{-0.24} \times 10^{22} \ \rm cm^{-2}$, which is similar
to the equatorial column density from the coupled \mytorus fit, and
similar to the radial column density from the spherical model fit.

The intrinsic continuum photon index is $1.502^{+0.007}_{-0.008}$,
similar to that obtained from the other two fits, and $f_{s} =
10.15^{+0.46}_{-0.31} \times 10^{-3}$, which is $\sim20\%$ lower than
the corresponding value from the coupled \mytorus fit.  The
temperature of the optically-thin emission, $kT_{\rm apec}$, is
statistically consistent with that obtained from both of the other
fits, but its normalization is $\sim19\%$ higher than that from the
coupled \mytorus fit.  The shift of the peak of the \fekalfa line is
statistically consistent with that obtained from the BN11 fit, but is
$\sim19\%$ higher than that from the coupled \mytorus fit. On the
other hand, the FWHM of the \fekalfa line, its flux, and EW, are all
statistically indistinguishable between the coupled and decoupled
\mytorus fits. Likewise, the centroid energies, fluxes, and EW values
of all eight Gaussian components are also statistically
indistinguishable between the coupled and decoupled \mytorus fits. The
observed continuum fluxes and luminosities are all consistent within
$1\%$ between the coupled and decoupled \mytorus fits, but the
intrinsic luminosities appear to be systematically higher for the
decoupled \mytorus fits, by $\sim6\%$ to $11\%$.

\begin{figure}
\centerline{
        \includegraphics[width=6cm,angle=270]{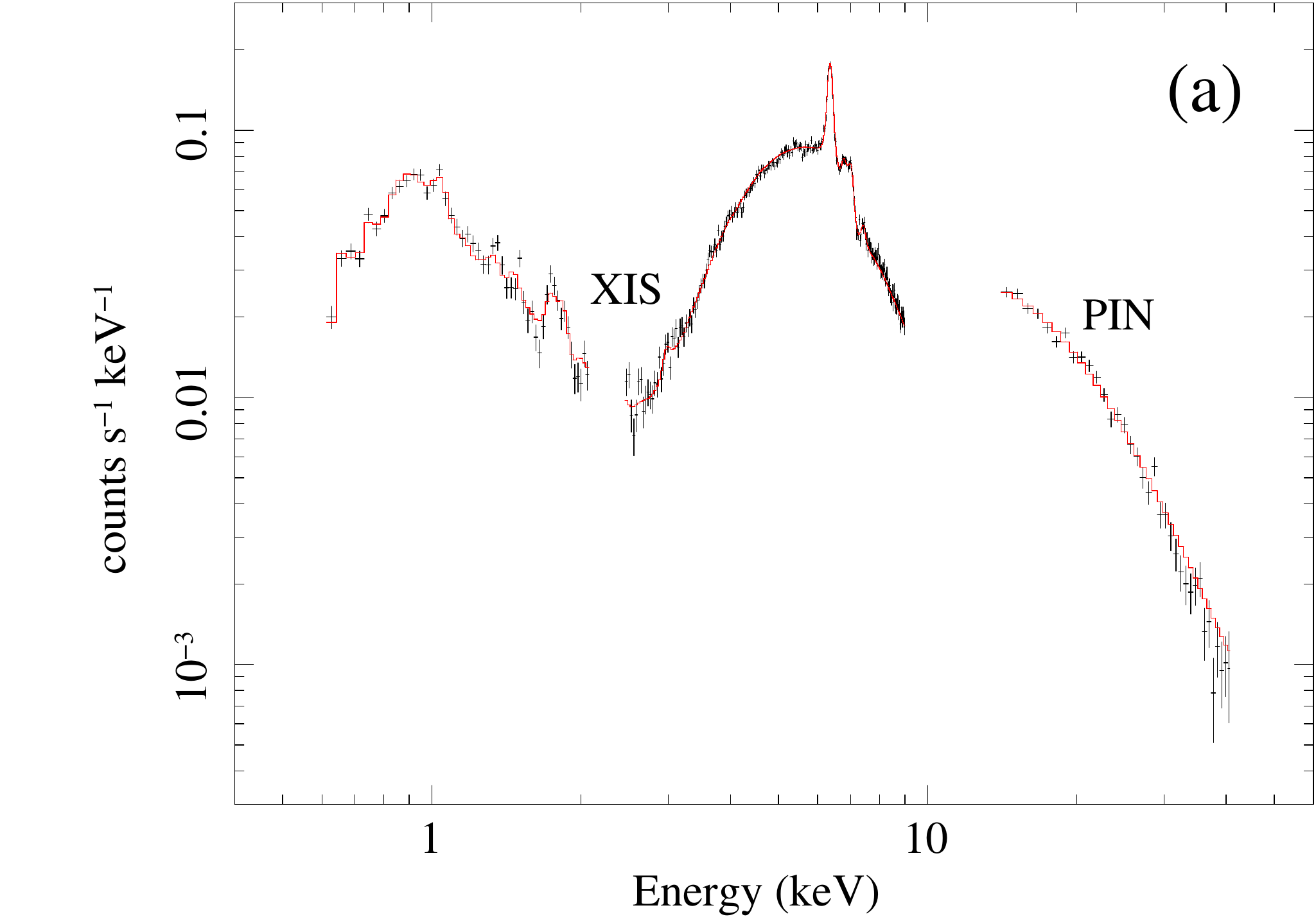}
        \includegraphics[width=6cm,angle=270]{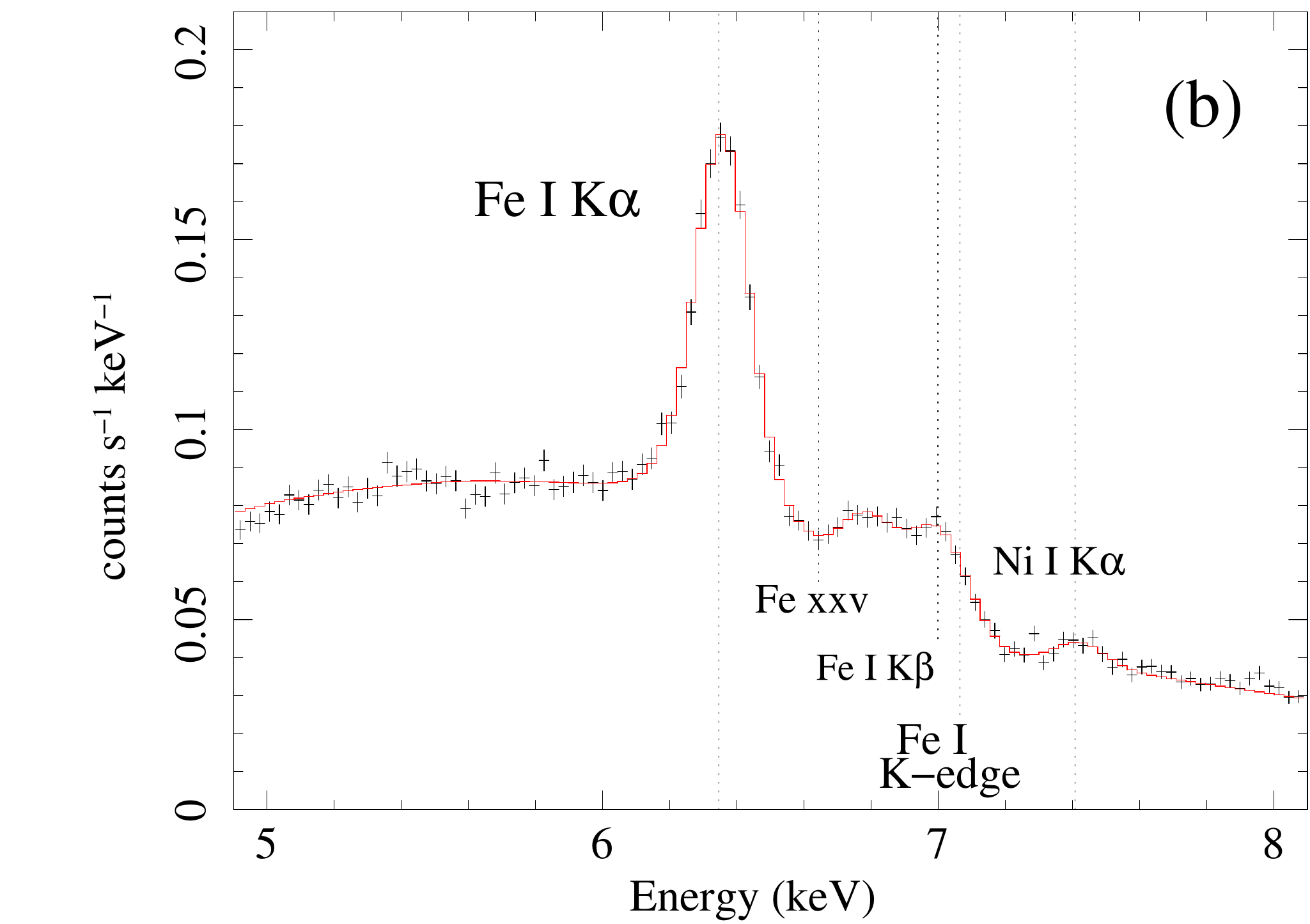}\hspace{1mm}
        }
\centerline{
        \includegraphics[width=3.43cm,angle=270]{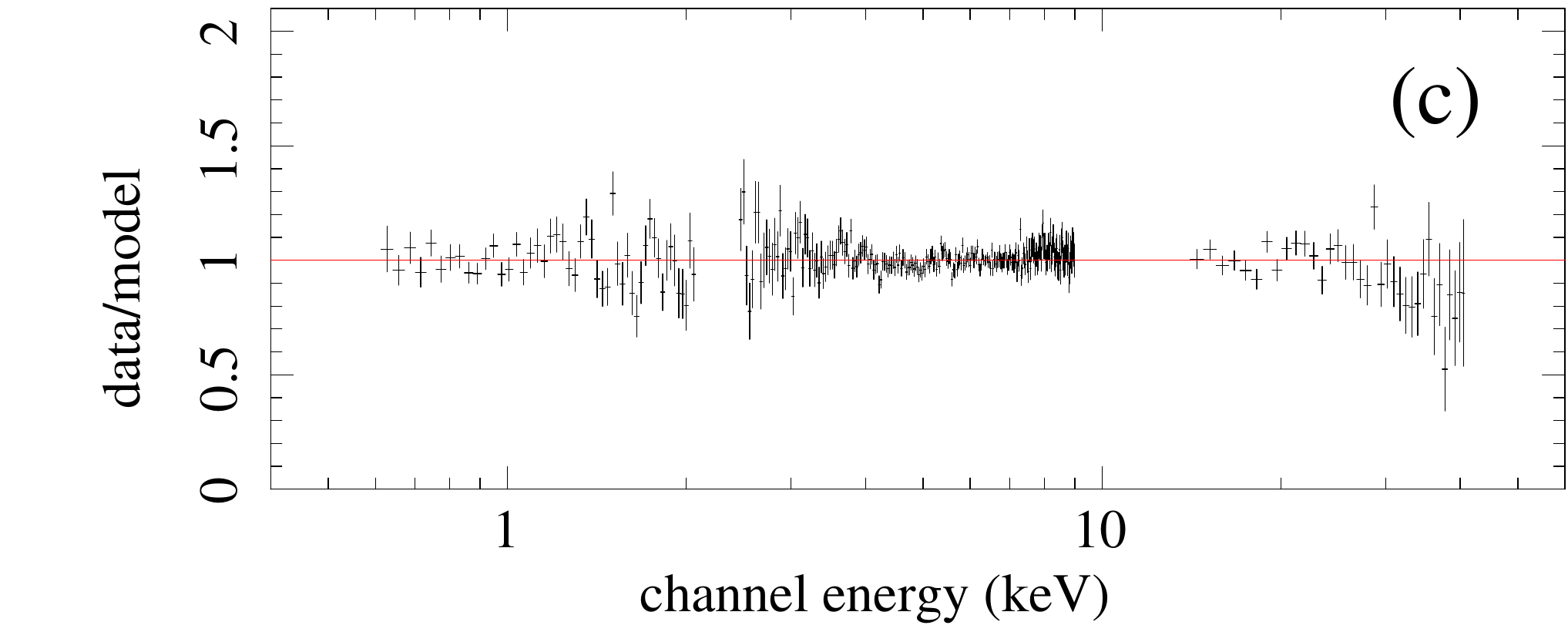}
        \includegraphics[width=3.43cm,angle=270]{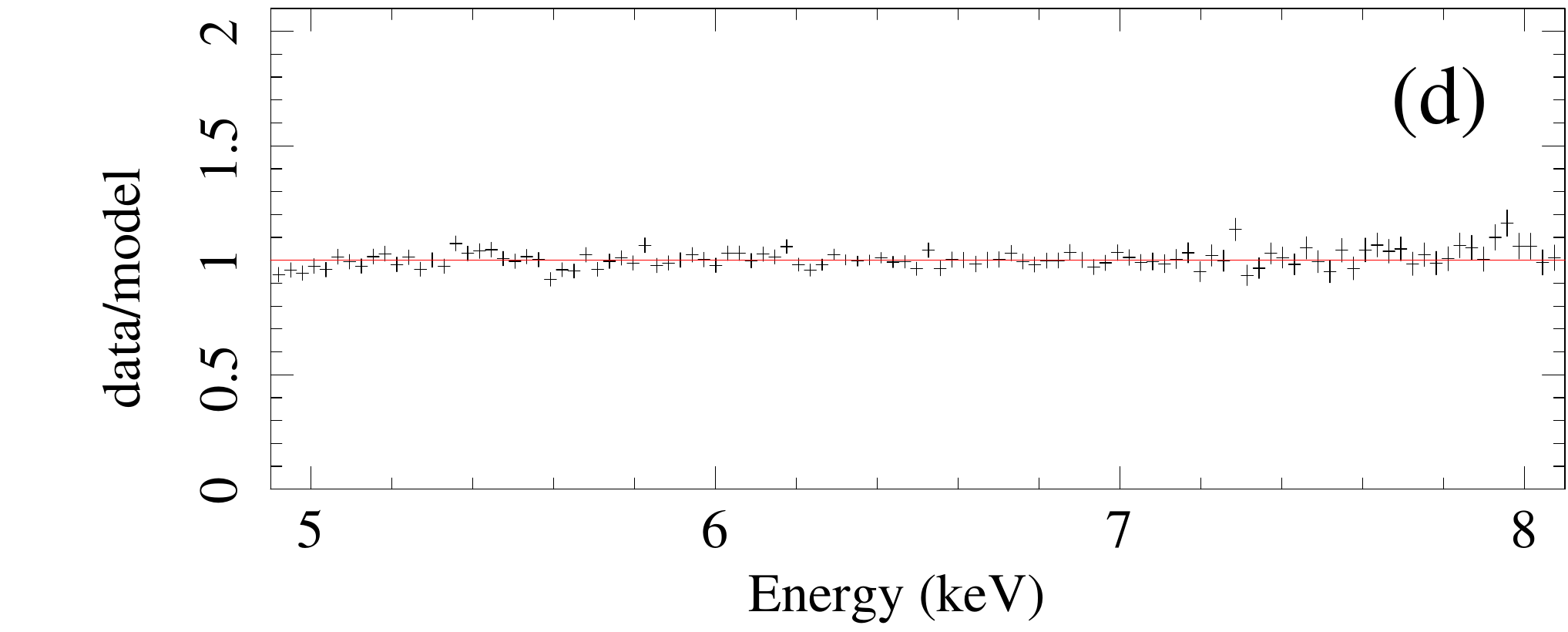}
        }
\caption{\small  Broadband spectral fit to the \src \suzaku spectra with the
decoupled \mytorus model (see \S\ref{decoupmytfits} and \tablespfitsp).
All of the descriptions in the caption for Fig.~\ref{fig:szsphdatrat} apply here.
}

\label{fig:szmytdecoupdatrat}
\end{figure}

\begin{figure}
\centerline{
        \includegraphics[width=6cm,angle=270]{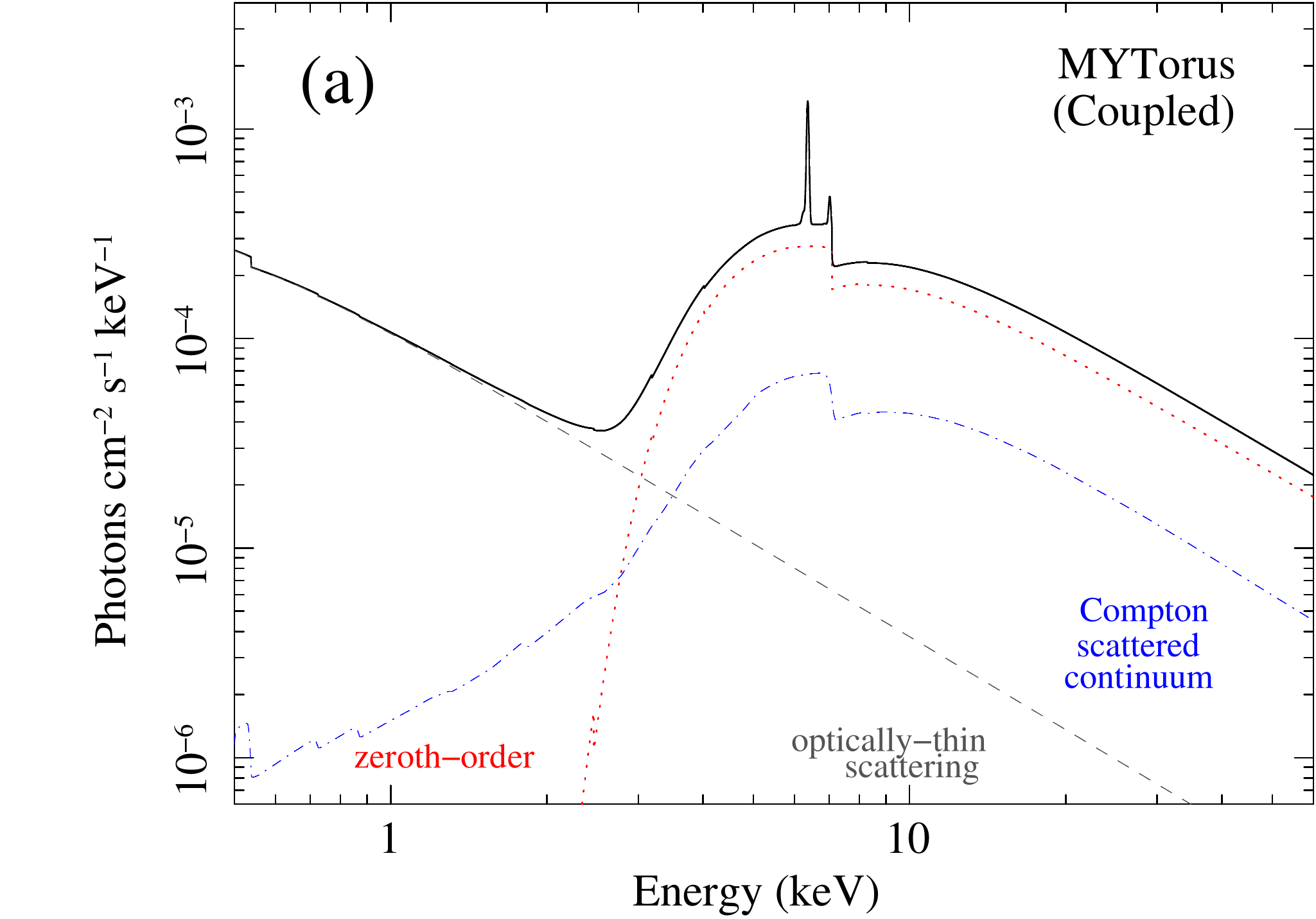}
        \includegraphics[width=6cm,angle=270]{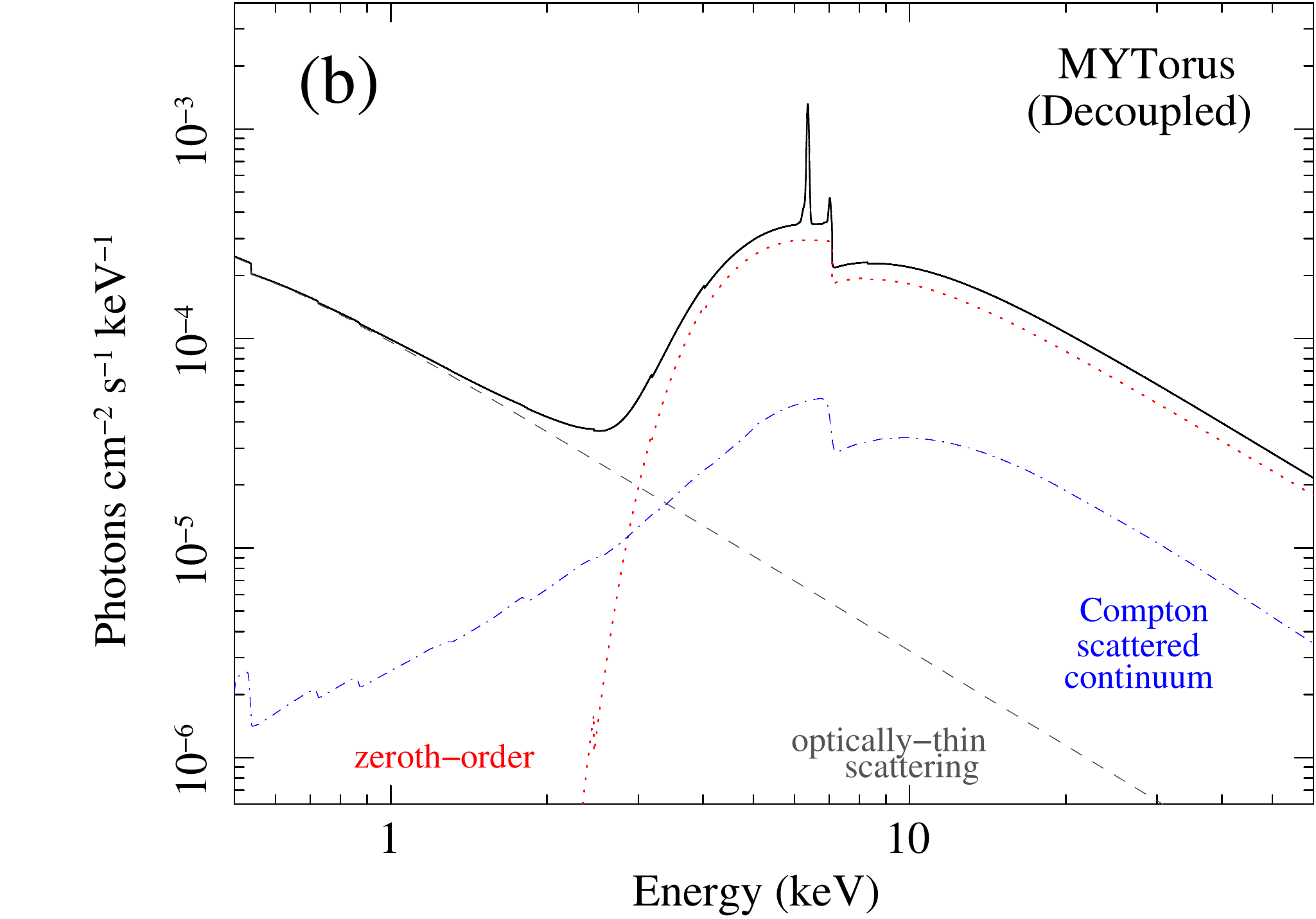}
        }
        \caption{\small A comparison of the principal model
          components for (a) the coupled \mytorus model, and (b) the
          decoupled \mytorus model (see \tablespfitsp). In each panel,
          the grey dashed line is the second power-law continuum, due
          to optically-thin scattering, the red dotted curve shows the
          zeroth-order continuum, the blue dot-dashed curve shows the
          \mytorus Compton-scattered continuum, and the solid black
          curve is the net spectrum. The latter includes the \fekalfa
          and \fekbeta emission lines. However, the individual
          emission-line components and the \apec thermal continuum are
          omitted for clarity (see text for details).  }
\label{fig:szcmpmytmodels}
\end{figure}

In Fig.~\ref{fig:szcmpmytmodels}, the best-fitting components of the
\mytorus coupled and decoupled models are directly compared with each
other. The red dotted curves show the zeroth-order continua, and the
blue dot-dashed curves show the Compton-scattered continua. The grey
dashed lines show the optically-thin scattered power-law continua.
The solid black curves show the sums of these separate components,
including the \fekalfa line emission. The separate contributions from
the \fekalfa line and the eight Gaussian components, as well as the
optically-thin thermal emission, are not shown for the sake of
clarity, since the purpose of the plot is to compare the \mytorus
continuum components.  It can be seen that the observed X-ray spectrum
of \src above $\sim4$~keV is dominated by the line-of-sight column
density, as opposed to the Compton-scattered continuum from the global
matter distribution. However, the latter is the continuum linked to
the observed \fekalfa emission line because it involves the same
global matter distribution as the line emitter. The solid angle
subtended by the line-of-sight material is negligble, so it cannot
produce any significant \fekalfa line emission.
Fig.~\ref{fig:szcmpmytmodels} also shows that the zeroth-order
spectral components for the coupled and decoupled \mytorus fits are
very similar, as are the Compton-scattered continuum components.

\section{Constraints on a Relativistic Disk Contribution to the X-ray Spectrum}
\label{relline}

\begin{figure}
\centerline{
        \includegraphics[width=7.2cm,angle=270]{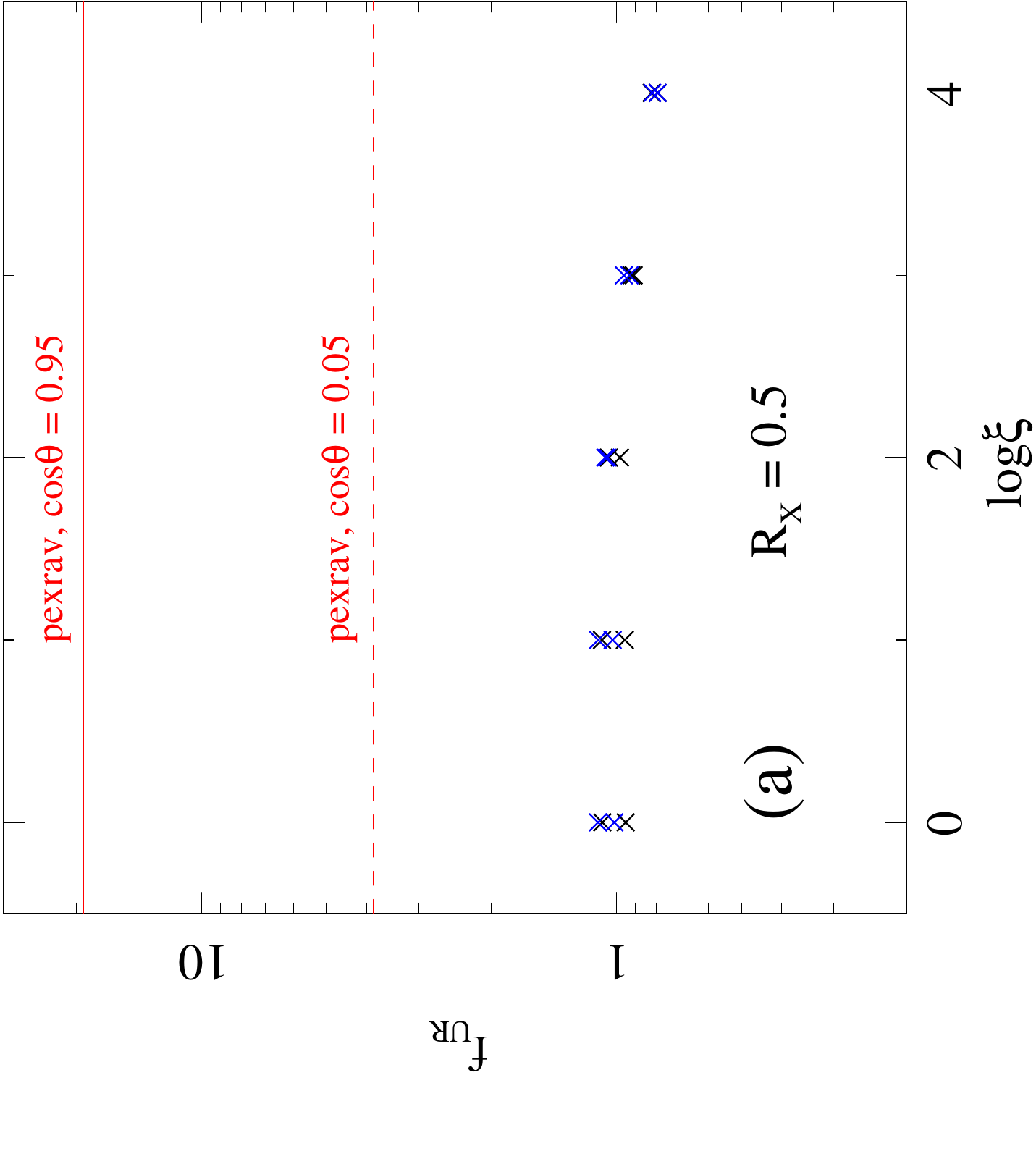}
        \includegraphics[width=7.2cm,angle=270]{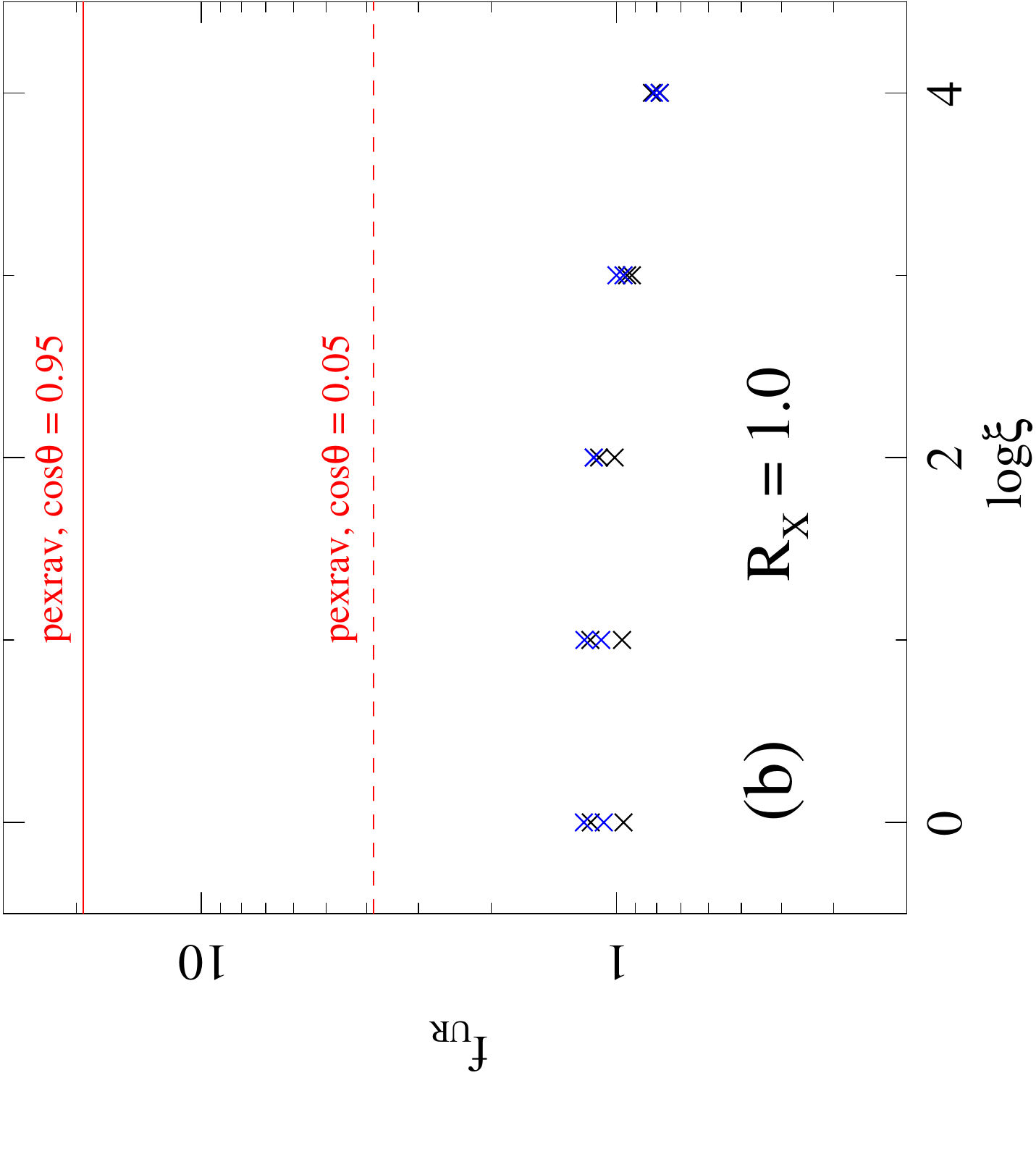}
        }
\centerline{
        \includegraphics[width=7.2cm,angle=270]{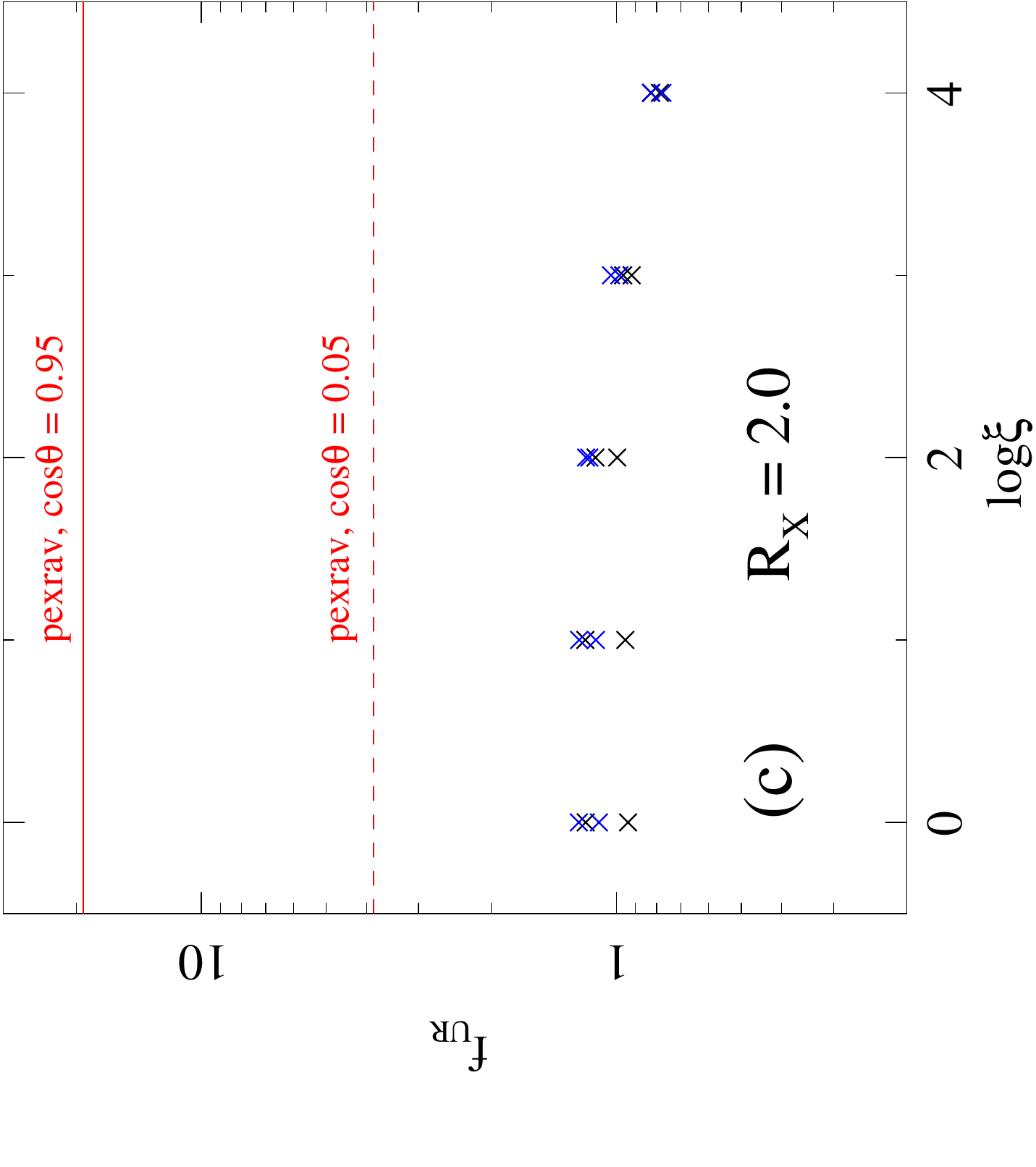} 
        \includegraphics[width=7.2cm,angle=270]{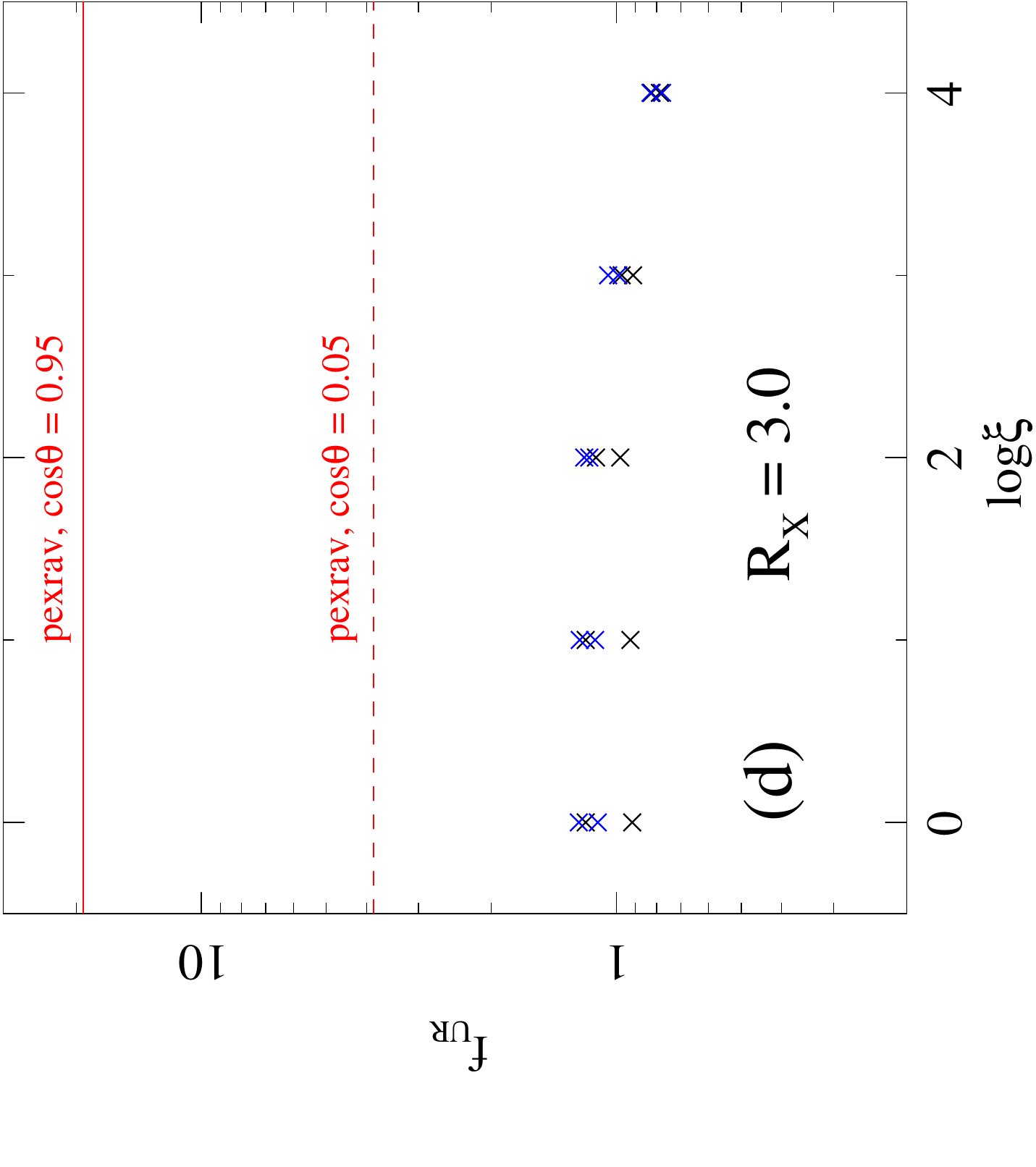}
        }
        \caption{\small The upper limits on the 2--20~keV flux contribution of the disk-reflection component, $f_{\rm UR}$, as a percentage of the total 2--20~keV flux (the latter includes the reflection and the uniform spherical model), versus the logarithm of the disk ionization parameter, $\log{\xi}$. The values of $f_{\rm UR}$ and $\log{\xi}$ are from \tablerelxillfitsp, and were obtained from fitting the \src \suzaku spectra with the uniform spherical model and the ionized disk-reflection model, \relxillp, as described in the text. Panels (a), (b), (c), and (d) show the results for the four values of the \relxill parameter, $R_{X}$, of 0.5, 1.0, 2.0, and 3.0, respectively. 
Black and blue crosses correspond to black-hole spin values of $a=0$ and $a=0.9982$ respectively. For clarity, the two inclination angles corresponding to $\cos{\theta_{\rm obs}}=0.05$ and $\cos{\theta_{\rm obs}}=0.95$ (see \tablerelxillfitsp), are not distinguished in the plots 
because there would be even greater overlap of the differnt symbols, due to the weak dependence of $f_{\rm UR}$ on the inclination angle.
As familiar benchmarks, in comparison, each panel also shows horizontal lines corresponding to the equivalent 2--20~keV flux percentage contributions for the \pexrav model, for the extremal inclination angles, $\theta$, corresponding to $\cos{\theta}=0.05$ (red, dashed lines), and $\cos{\theta}=0.95$ (red, solid lines).
} 
\label{fig:relxillulone}
\end{figure}

\begin{figure}
\centerline{
        \includegraphics[width=6.5cm,angle=270]{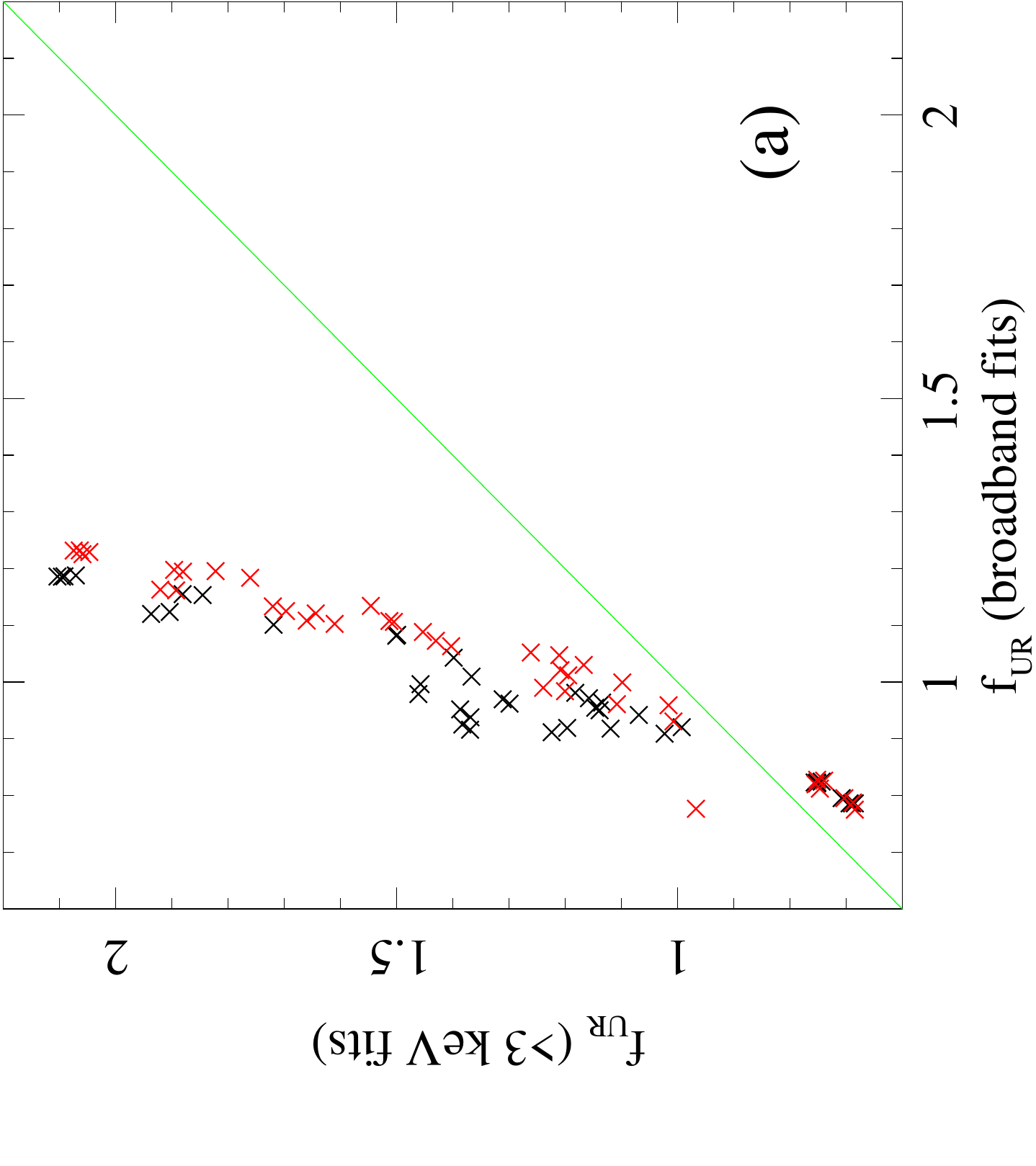}
        \includegraphics[width=6.5cm,angle=270]{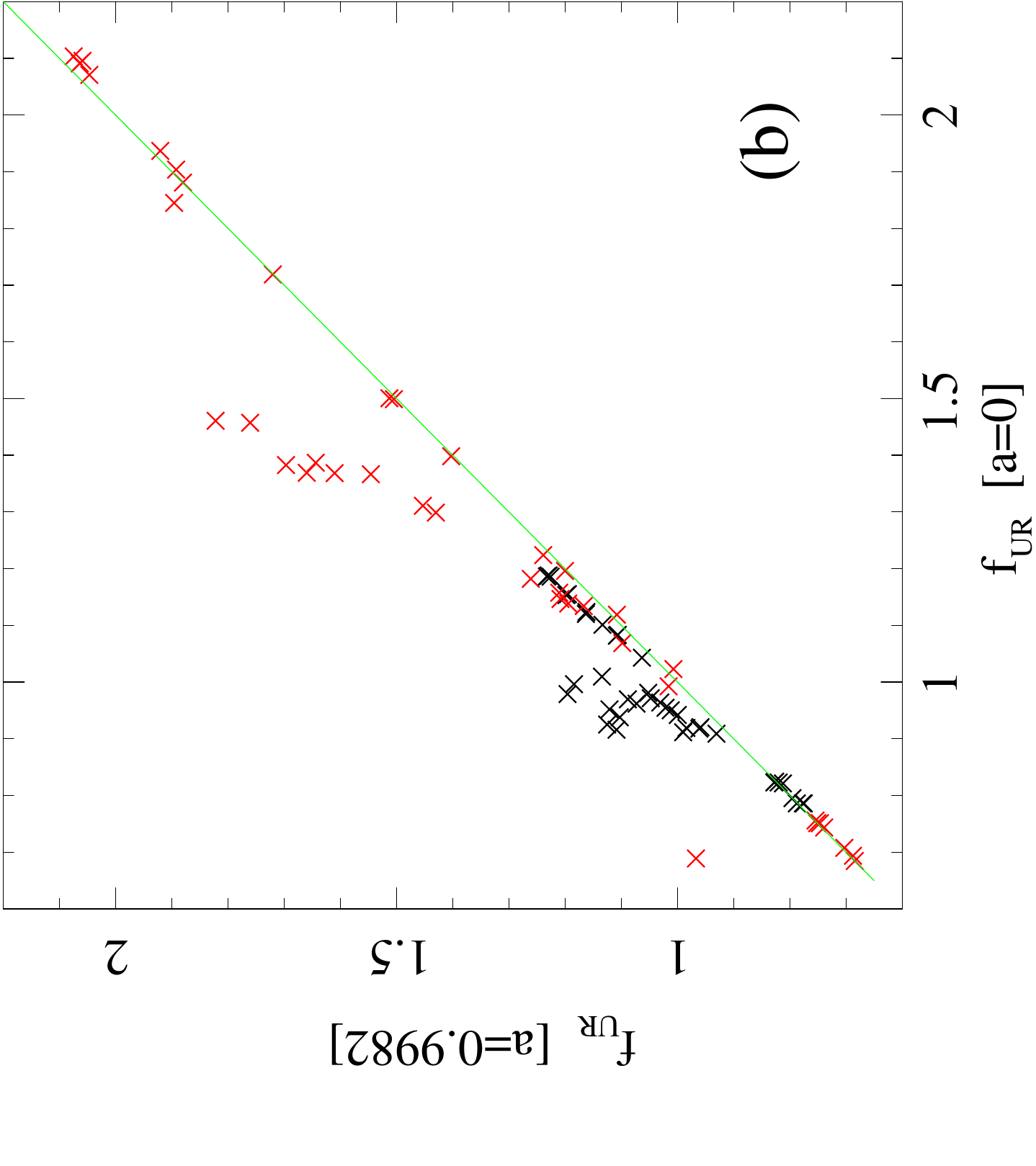}
        }
        \caption{\small (a) From \tablerelxillfitsp, the values
          of the percentage contributions of the disk-reflection model
          component, $f_{\rm UR}$, obtained from the high-energy fits
          above 3 keV, versus the corresponding values obtained from
          the broadband fits. (b) The effect of the black-hole spin
          parameter, $a$, on the values of $f_{\rm UR}$. Shown are the
          percentage contributions of the disk-reflection model
          component, $f_{\rm UR}$, in \tablerelxillfitsp, for the
          black-hole spin parameter, $a$, equal to the extremal upper
          value of 0.9982, versus the values of $f_{\rm UR}$ obtained
          for $a$ equal to the extremal lower value of 0. In each
          panel, the green line is the locus of points that have equal
          values of $f_{\rm UR}$.  }
\label{fig:relxillultwo}
\end{figure}

We have shown that the \suzaku nuclear X-ray spectrum of \src above
$\sim2$~keV can be explained with a model that is as simple as a
uniform, spherically-symmetric, solar-abundance matter distribution
that is more than $10^{4}$ gravitational radii from the putative
central black hole.  The data/model residuals in the Fe~K band after
fitting any of the nonrelativistic models described thus far, are such
that adding a relativistic disk model to such fits would simply be
fitting noise (see Figs.~\ref{fig:szsphdatrat},
\ref{fig:szmytcoupdatrat}, and \ref{fig:szmytdecoupdatrat}). This
implies that there is no requirement for a relativistically-broadened
\fekalfa emission line.  Nevertheless, we can quantify upper limits on
the relative contribution of a reflection spectrum from a relativistic
accretion disk. For the latter we use the ionized, relativistic
disk-reflection model \relxillp, which is described in detail in
Garc\'{i}a \etal (2014), and commonly applied to the X-ray spectra of
accreting sources. We add this model to the uniform, spherical (BN11)
model of distant matter described in \S\ref{sphericalfit}. Since the
best-fitting BN11 model has a marginally higher fit statistic than the
two \mytorus model fits, using the BN11 model as the baseline for the
relativistic disk fit will provide a more stringent test for the
presence of any disk component in the X-ray spectrum.  In addition,
the BN11 model allows us to tie the Fe abundance in the distant
matter, to that in the disk.  Generally, we are addressing the
question of why there appear to be no signatures of the accretion disk
in the X-ray spectrum of \srcp, and therefore, we are searching for
scenarios in which the upper limit on the disk contribution allowed by
the data is as large as possible.

The \relxill model has 14 parameters, although one of these is the
cosmological redshift, which is fixed at the value given in
\S\ref{intro}.  Since the BN11 model fit leaves no margin for any
significant spectral complexity (especially in the Fe~K band), the
\relxill parameters cannot be constrained if too many of them are
free. Therefore, we take the approach of selecting a wide range of
combinations of the \relxill parameters, freezing all except the
overall normalization of the \relxill component, and studying the
upper limits on that normalization as a function of the unique sets of
combinations of the other parameters. Specifically, for each unique
parameter set, we vary the \relxill overall normalization until the
\chisq value relative to the BN11 baseline fit increases by 2.706,
corresponding to 90\% confidence for one free parameter. We now
briefly describe each of the other 12 parameters, and their treatment
for the study.

{\it Disk radial emissivity parameters}: $q_{1}$, $q_{2}$, $r_{\rm
  BR}$. The radial emissivity in the \relxill model is described by a
broken power law, with power-law indices $q_{1}$ and $q_{2}$ for radii
$\le r_{\rm BR}$ and $> r_{\rm BR}$ respectively ($r_{\rm BR}$ is
simply the ``break radius'' at which the power-law index switches
between $q_{1}$ and $q_{2}$). Observationally, the radial emissivity
is poorly constrained in AGN in general, and from a theoretical
point-of-view, there is a wide range of possibilities.  It is
challenging to constrain the slope of even a single power-law
emissivity parameterization for most AGN, and this is true of \src as
well, so it has to be fixed.  In our fits we use a single power-law
function for the radial emissivity by tying $q_{1}$ and $q_{2}$
together and fixing the index at a value of 3.0 (the value of $r_{\rm
  BR}$ is then irrelevant, but we fix it at $15r_{g}$). The value of
3.0 for $q_{1}$ ($=q_{2}$) is that expected for the outer parts of a
Shakura and Sunyaev disk (e.g., Laor 1991; Dauser \etal 2013). Studies
of samples of AGN find that emissivity indices do not often go below
2.0, and emissivity indices greater than 3.0 are required by some AGN
(e.g., Walton \etal 2013). A steeper radial emissivity law would
produce a stronger reflection spectrum from the inner disk, and
therefore, {\it smaller} upper limits on the \relxill component
contribution to the net X-ray spectrum. The present study aims to
establish the largest possible contributions of the disk-reflection
spectrum, so an emissivity index of 3.0 is appropriate.

{\it Intrinsic continuum photon index}: $\Gamma$. The intrinsic
continuum in the \relxill model is a power law with an exponential
high-energy cut-off. The photon index, $\Gamma$, is tied to the
corresponding photon index of the BN11 model.

{\it Cut-off energy}: $E_{\rm cut}$. This is fixed at 300~keV. It is
beyond the bandpass of the data and does not affect the
fit. Nevertheless, it is consistent with the findings from studies of
samples of AGN observed by \bsax (Dadina 2007, 2008), and \nustar
(Balokovi\'{c} \etal 2020).

{\it Iron abundance relative to solar}: $X_{\rm Fe}$. This is tied to
the relative iron abundance in the BN11 model.

{\it Black-hole spin}: $a$. The two extremal values for this
dimensionless black-hole spin parameter are selected, namely 0.0, and
0.9982, the latter corresponding to the maximal value for an accreting
black hole (Thorne 1974).

{\it Inner disk radius}: $R_{\rm in}$. This parameter is set to -1,
which means that the \relxill model will use a value for the inner
disk radius that corresponds to the inner-most stable orbit for the
value of black-hole spin selected. This is aligned with our goal of
obtaining the largest upper limits on a disk-reflection contribution
to the X-ray spectrum of \srcp, because we are investigating what
contribution the data allow, given that the reflection appears to be
absent.  Setting $R_{\rm in}$ to larger values (and thereby emulating
a disk with a truncated inner edge), would make the \fekalfa
emission-line narrower, which would in turn {\it reduce} the upper
limit on a disk-reflection contribution. This is because the \fekalfa
line in the \src data is already fully accounted for by the
distant-matter contribution, and it is locked in by the other spectral
signatures of the distant matter.

{\it Outer disk radius}: $R_{\rm out}$. The outer disk radius is fixed
at $1000r_{g}$, the maximum value in the the model table. Relativistic
features in the spectrum are formed within tens of gravitational
radii, and if the outer radius were confined to such small values,
smaller upper limits would be obtained on the \relxill
normalization. Therefore, the choice of $1000r_{g}$ will give the
largest upper limits on the latter.

{\it Disk inclination angle}: $\theta_{\rm obs}$. Valid inclination
angles of the disk in the \relxill model lie in the range $0.95 \ge
\cos{\theta_{\rm obs}} \ge 0.05$, and we select the two extremal
values, corresponding to $18.2^{\circ}$ and $87.1^{\circ}$
respectively.

{\it Logarithm of the ionization parameter}: $\log{\xi}$. The
ionization parameter is essentially the ratio of incident ionizing
flux on the disk to the ion number density.  We select five values of
$\log{\xi}$: 0.0, 1.0, 2.0, 3.0, and 4.0. The lowest value corresponds
to an essentially neutral disk, and the highest value corresponds to a
completely ionized disk.

{\it Reflection fraction}: $R_{X}$. This is not the same as the
relative reflection strength in standard nonrelativistic disk models
such as \pexrav (Magdziarz and Zdziarski, 1995). Dauser \etal (2016)
explain in great detail the meaning of $R_{X}$ in the \relxill model,
and how it compares with other models and definitions of the relative
magnitude of the reflection spectrum compared to the intrinsic
continuum (see also Dauser \etal 2014).  For \relxillp, $R_{X}$ is
essentially the ratio of the coronal intensity that illuminates the
disk to the coronal intensity that reaches the observer. Unlike other
definitions of the reflection fraction, $R_{X}$ does not depend on the
inclination angle, nor the physical condition of the disk. We select
four values for $R_{X}$, namely, 0.5, 1.0, 2.0, and 3.0. Considering
that the \src data do not require any reflection, this range more than
covers the scenarios of interest (see Dauser \etal 2016).

Thus, there are 80 unique combinations of parameters: two values of
black-hole spin, two values of disk inclination angle, four values of
$R_{X}$, and five values of \logxip.  This will lead to 80
corresponding upper limits on the contribution of the \relxill
component to the overall X-ray spectrum of \srcp. This number of
parameter combinations avoids an overwhelming number of spectral fits,
yet spans a wide enough parameter space to cover a plausible range of
scenarios. In each of the 80 spectral fits, all of the BN11 parameters
that were free in the analysis with only the BN11 model
(\S\ref{sphericalfit}) remain free, and in the \relxill component,
only the \relxill overall normalization is free. Each of the fits
begins with a best-fitting value of the \relxill normalization of 0.0,
and the upper limit of that normalization is obtained as described
above.

The \relxill normalization parameter by itself is not very
informative. Instead, we introduce a quantity derived from it, $f_{\rm
  UR}$, defined as the upper limit on the percentage of the 2--20~keV
net flux that is in the \relxill component, where the net flux
includes the BN11 and \relxill components. Thus, $f_{\rm UR}$ is a
measure of the maximum percentage contribution of the relativistic
disk component to the total 2--20~keV spectrum that is allowed by the
data (at the 90\% confidence level), for a given set of the other
\relxill parameters, aside from its normalization. In order to place
values of $f_{\rm UR}$ in a familiar context, we can calculate an
analogous 2--20~keV flux contribution for the commonly used
nonrelativistic disk-reflection model, \pexravp, which can be directly
compared with $f_{\rm UR}$ obtained from the \src fits with
\relxillp. We calculate that flux percentage for \pexrav for the
extremal values of the disk inclination angle, $18.2^{\circ}$ and
$87.1^{\circ}$, both for a \pexrav reflection fraction, $R$, of 1.0.
For this calculation, we set the photon index of the intrinsic
power-law continuum, $\Gamma$, to the best-fitting value found from
the BN11 spherical model fit (\tablespfitsp). We will refer to these
two benchmark \pexrav flux percentages as $f_{\rm PB1}$ and $f_{\rm
  PB2}$ for disk inclination angles $18.2^{\circ}$ and $87.1^{\circ}$,
respectively.  The values of $f_{\rm PB1}$ and $f_{\rm PB2}$ are
calculated to be $3.845\%$ and $19.238\%$ respectively.  Thus, near
face-on and edge-on nonrelativistic, neutral, reflection contributes
less than $\sim4\%$ and $\sim20\%$ to the net 2--20~keV flux,
respectively. Note that the \pexrav model does not include any
fluorescent line emission.

The 80 values of $f_{\rm UR}$ that we obtained from the 80
configurations of the BN11 and \relxill models applied to \src are
shown in \tablerelxillfitsp. The same 80 values of $f_{\rm UR}$ are
shown in Fig.~\ref{fig:relxillulone}, plotted against \logxip,
separated into four panels by the value of $R_{X}$, as indicated in
each panel. For direct comparison, each panel of
Fig.~\ref{fig:relxillulone} also shows the two \pexrav benchmarks,
$f_{\rm PB1}$ and $f_{\rm PB2}$. It can be seen from \tablerelxillfits
and Fig.~\ref{fig:relxillulone}, that the values of $f_{\rm UR}$ are
clustered within the range $\sim1\pm 0.25\%$ across the entire
parameter space that is probed. The values of $f_{\rm UR}$ appear to
be insensitive to $R_{X}$, but are systematically lower for the
largest value of the ionization parameter. It can also be seen that
all values of $f_{\rm UR}$ are a factor of $\sim3$ smaller than the
lowest \pexrav benchmark, and nearly a factor 20 smaller than the
highest \pexrav benchmark.

It could be argued that it is the soft part of the X-ray spectrum that
is constraining the disk-reflection spectrum to be negligible, as
opposed to the lack of a relativistically broadened \fekalfa
emission-line component.  In order to address this, we repeated the
entire analysis described above, using only the spectral data above
3~keV. Since $f_{\rm UR}$ is defined by fluxes in the 2--20~keV band,
models fitted above 3~keV were extrapolated down to 2~keV in order to
calculate $f_{\rm UR}$. This approximation has a negligible effect on
$f_{\rm UR}$, compared to the differences in the $f_{\rm UR}$ values
between the broadband and higher energy fits.  The numerical results
for $f_{\rm UR}$ are shown in columns 6 and 7 in \tablerelxillfitsp,
and Fig.~\ref{fig:relxillultwo}(a) shows the values $f_{\rm UR}$
obtained from the fits above 3~keV, plotted against the corresponding
values from the broadband fits. It can be seen that the values of
$f_{\rm UR}$ obtained from the fits above 3~keV are indeed
systematically higher than those obtained from the broadband fits,
where $f_{\rm UR}$ for the latter is $<1.9\%$. However, even in the
worst case, for the fits above 3~keV, $f_{\rm UR}<2.1\%$, which is
still nearly an order of magnitude less than the face-on \pexrav
benchmark value (see Fig.~\ref{fig:relxillulone}).

Fig.~\ref{fig:relxillultwo}(a) also shows that there is a relatively
narrow range in $f_{\rm UR}$, regardless of the values of any of the
\relxill parameters that were varied.  The black and red data points
in Fig.~\ref{fig:relxillultwo}(a) correspond to fits with the
black-hole spin equal to 0.0 and the maximum, 0.9982, respectively,
and it can be seen that the fits above 3~keV tend to have slightly
higher values of $f_{\rm UR}$ for a=0.0 than for
$a=0.9982$. Fig.~\ref{fig:relxillultwo}(b) shows the dependence on
black-hole spin in a different way, in which values of $f_{\rm UR}$
for the maximum spin are plotted against corresponding values for the
minimum spin. Values of $f_{\rm UR}$ from the broadband fits are shown
in black, whereas those from the fits above 3~keV are shown in red. It
can be seen that for most cases, the black-hole spin does not affect
$f_{\rm UR}$, but there are clusters of data points for which the
maximum spin gives values of $f_{\rm UR}$ that are higher than the
corresponding values for $a=0$. However, the difference is never more
than $30\%$.

\section{Conclusions}
\label{conclusions}

We have presented the results of a new analysis of the \suzaku X-ray
spectrum of the Compton-thin Seyfert 2 galaxy \srcp.  We have applied
new models and methodologies that did not exist at the time that
results from this observation were first published, and have not been
applied in several subsequent analyses of the same data (see
appendix). Our study focusses on features above $\sim2$~keV, in order
to investigate whether a relativistically-broadened \fekalfa line and
associated Compton-reflection continuum, originating in an accretion
disk, are present in the X-ray spectrum. With modern techniques, these
broad features that are commonly reported in type~1 AGN, should just
as easily be detectable in Compton-thin type~2 AGN. In order to
achieve this, features from distant matter (thousands of gravitational
radii or more from the central black hole), namely the narrow \fekalfa
line and its associated Compton-reflection continuum, should be
modelled self-consistently, along with any line-of-sight absorption.

We applied three models for the neutral distant-matter components of
the spectrum: a uniform, spherical distribution of matter (using the
BN11 model of Brightman and Nandra, 2011), and two variations of the
\mytorus reprocessor model (Murphy and Yaqoob, 2009).  We found that
the uniform spherical model provided an excellent fit to the data,
yielding an Fe abundance that is $1.102^{+0.024}_{-0.021}$ relative to
solar.  The best-fitting radial column density for this model is
$2.58\pm0.02 \times 10^{23} \ {\rm cm}^{-2}$.  The simple model
simultaneously accounts for the flux and equivalent width of the
\fekalfa line, the associated Compton-reflection continuum, the
detailed spectral shape in the Fe-K edge region, and line-of-sight
absorption. These features fit well without any \adhoc adjustments to
the model, aside from additional \nika line emission (likely due to a
Ni overabundance), and an absorption line at $\sim6.7$~keV due to
He-like Fe resonance absorption in an ionized wind, possibly
associated with the accretion disk.  The uniform spherical model is
not a unique description of the data. We also fitted two variations of
the \mytorus model, both having a fixed solar Fe abundance, and found
that these models could not be ruled out.  Again, the models
simultaneously account for the key features in the high-energy
spectrum, without phenomenological adjustments.

The residuals between the data and all three of the distant-matter
models are essentially noise, so that there is no need for any
relativistic disk-reflection component.  We proceeded to quantify this
by taking the spherical model fit as a baseline, and systematically
finding upper limits on the contribution of a disk-reflection
component, in the form of the \relxill model, for a wide range in
parameter space. We found that the upper limits on the contribution of
such a component are not significant, and are remarkably insensitive
to the model parameters. Formally, in the 2--20~keV band, the
disk-reflection component never contributes more than $\sim10\%$ and
$\sim25\%$ of standard \pexrav reflection continua with a reflection
fraction of 1.0, for the most face-on and edge-on orientations
respectively. Given that a uniform spherical distribution is a viable
configuration of the X-ray obscuring matter in \srcp, we cannot
conclude that the orientation of the accretion disk relative to the
observer is edge-on or nearly edge-on. Indeed, there is no evidence
for supporting any particular orientation of the accretion disk in
\srcp.

Despite the fact that most historical X-ray spectral studies of \src
are based on phenomenological models that lack self-consistency, our
tight upper limits on the relativistic disk-reflection component are
in broad agreement with those obtained from the simpler historical
modelling. All except the very first \suzaku study (Shirai \etal 2008)
found that a reflection continuum was either relatively weak or
negligbile (see appendix for details, and references therein).
However, since we have shown that the line-of-sight absorption, the
\fekalfa and \fekbeta emission lines, Compton-scattered continuum, and
detailed Fe~K edge structure from that obscuring material are all
tightly locked together, there are two very important
implications. One is that a disk with a truncated inner edge cannot
explain the weak disk-reflection component. This is because truncation
makes the \fekalfa narrower, but the strength of the \fekalfa line is
already locked in by the other properties of the material that
produces the line. Therefore, a truncated disk would give even smaller
upper limits on the contribution of a disk-reflection component to the
net spectrum. The second implication is that a very high disk
ionization cannot explain the lack of relativistically-broadened
\fekalfa line.  Our analysis included completely ionized states of the
disk, yet the upper limits on the disk contribution to the net
spectrum were still small. This is because the shape of the disk
spectrum across the energy bandpass is sufficiently different to that
required by the data, and the self-consistent spectral components from
the obscuring material cannot be perturbed significantly by the
incongruent disk spectrum. Both a truncated disk and a highly ionized
disk are commonly invoked whenever a disk-reflection component is
absent when one was expected (Reynolds \etal 2021, and references
therein).

A simple spherical X-ray reprocessor has never been fitted to the
X-ray spectrum of \src from any mission.  In general, our analysis
highlights the need to carefully consider which models may be most
appropriate to apply to a particular source, and whether complex
models are uniquely required to describe the data better than simpler
ones. If relativistically-broadened \fekalfa line emission and
reflection are ubiquitous features in type~1 AGN, they should also be
ubiquitous in Compton-thin AGN, because the tools for deconvolving the
nonrelativstic features have improved since the early days of CCD
X-ray spectroscopy. Traditionally, it has been thought that type~1 AGN
are better for studying relativistically-broadened \fekalfa lines and
the X-ray reflection continuum. However, somewhat counterintuitively,
Compton-thin AGN can be better in some cases, because the
line-of-sight extinction, signatures from the global matter
distribution, and the strong narrow \fekalfa emission line are
strongly locked together. When these are fitted with self-consistent
models, it results in better constraints on any broad-line component,
whereas type~1 AGN have a narrow \fekalfa that cannot be anchored by
associated absorption and scattering features. Moreover, in
Compton-thin AGN, the very complex soft X-ray portion (below
$\sim2$--4~keV) of the disk-reflection spectrum is suppressed, because
it is absorbed by more distant matter, considerably facilitating the
spectral deconvolution problem for disentangling the disk and
distant-matter features.

More Compton-thin AGN should be re-examined along the lines that we
have presented for \src (see also Tzanavaris \etal 2021), to test
whether they systematically have a dearth of relativistic signatures
in the X-ray spectra. Two of the most common explanations do not work
for \srcp.  In addition to the specific reasons given for \src above,
in general, it would be problematic if Compton-thin AGN were
universally required to have truncated and/or extremely ionized disks,
because current unification models of type~1 and type~2 AGN insist
that there is no difference in the central engine for the
classification of AGN type. We have also noted that the Fe abundance
in distant matter cannot be wildly different to the Fe abundance in
the accretion disk in the same AGN. This is another advantage of using
Compton-thin AGN to study or search for relativistic disk signatures,
because the Fe abundance can be strongly constrained by the multiple
features from distant matter, and it can be tied to the disk Fe
abundance for spectral fitting. However, as far as we are aware, this
fact has never been utilized for studies reported in the current
literature. Moreover, a very high Fe over-abundance is often required
as an unavoidable price to pay for accounting for an observed broad
\fekalfa line in an AGN or X-ray binary system. There is no evidence
for supersolar abundances in other parts of these systems or the host
galaxt, yet an Fe overabundance as high as a factor of $\sim8$ in the
accretion disk is not considered unusual in these studies (e.g., see
Reynolds 2021, and references therein). In one case, overabundances of
10 to 20 (for different models), were invoked for a particular AGN
(see Garc\'{i}a \etal 2018, and references therein).  However, none of
these studies has ever applied self-consistent models of the
distant-matter spectrum to constrain the Fe abundance in the disk. In
fact, the Fe abundance in the distant matter and in the accretion disk
{\it cannot} be tied together if the distant-matter model components
are empirical and/or \adhocp.  The case of \src should prompt
theoretical studies that examine the current paradigms more
critically.

The considerations discussed above will be all the more important for
forthcoming \xrism data for AGN in general, because \xrism has no hard
X-ray detector with high throughput coverage above 12~keV, so a
critical portion of the Compton-reflection continuum will be
missing. Only a limited number of observations are likely to have
simultaneous coverage with \nustarp. On the other hand, the superior
\xrism spectral resolution in the Fe~K band will improve the ability
to deconvolve the distant-matter features. The Fe abundance will play
a critical role in these studies, and the greater sensitivity of the
\xrism calorimeter to narrow \nika line emission will elevate the
importance of the role of the Ni abundance as well (e.g., see Yaqoob
and Murphy 2011; Fukazawa \etal 2016).

\section*{Acknowledgements}
The authors thank T. J. Turner, A. Scholtes, M. Trevor, and
M. M. Tatum for their contribution to this work.  This work was
supported by NASA grants NNX10AE83G and 80GSFC21M0002. P.T. also
acknowledges support from NASA grant 80NSSC18K0408 (solicitation
NNH17ZDA001N-ADAP).  This research has made use of data and software
provided by the High Energy Astrophysics Science Archive Research
Center (HEASARC), which is a service of the Astrophysics Science
Division at NASA/GSFC and the High Energy Astrophysics Division of the
Smithsonian Astrophysical Observatory.

\section*{Data Availability}
The data underlying this article will be shared on reasonable request
to the corresponding author.

\bsp

\appendix
\section{Comparison with Previous X-ray Spectral Studies of NGC~4388}
\label{appendixa}
\setcounter{table}{0}
\renewcommand{\thetable}{\Alph{section}\arabic{table}}

In this section we summarize results from previous studies of \srcp,
from observations with \suzaku and other X-ray astronomy
missions. Since there are a large number of observations, rather than
sequentially describe results from every mission, we give the
principal results organized by particular spectral features or
characteristics. We focus only on those results that are relevant for
placing our results for the \suzaku spectral data into context. In
particular, we do not go into great detail about the soft X-ray
spectrum (below $\sim2$~keV), because the present paper is concerned
with the \fekalfa line and Compton-reflection continua. Some
comparisons of our results with historical ones are given in the main
body of the present paper and are not necessarily repeated here.

\tablehistory shows the principal X-ray missions that have observed
\src since \ascap, and the associated principal references for studies
that have been published using the data.  Note that the extended
\fekalfa emission in the \chandra data reported by Yi \etal (2021) has
less than 2 per cent of the flux of the \fekalfa line from the nuclear
region. This means that in data that are not spatially resolved (i.e.,
all other data represented in \tablehistoryp), the extended component
makes a negligible contribution to the observed \fekalfa line. Note
that the Shu \etal (2011) study is based in \chandra grating data, so
also does not spatially resolve the \fekalfa line. Also note that some
of the studies that are cited use data from more than one mission, so
some of the references in \tablehistory are repeated.

After Shirai \etal (2008), results for the same \suzaku data were also
published by Miyazawa \etal (2009), Lubi\'{n}sky \etal (2016),
Fukazawa \etal (2011, 2016), and Kawamuro \etal (2019). The study by
Fukazawa \etal (2016) was entirely focussed on the \fekalfa and \nika
emission-line features, and the analysis was resticted to the 5--9~keV
energy band.  The work by Lubi\'{n}sky \etal (2016) combined
non-simultaneous INTEGRAL, \xmmp, and \suzaku spectra, so the model
fits are compromised because of well-known spectral and flux
variability. Also, although they included a Gaussian component for the
narrow \fekalfa line, they did not give any spectral-fitting results
for the line. Therefore, the Lubi\'{n}sky \etal (2016) results will
not be mentioned further.

\subsection*{Flux and Luminosity} 
The continuum flux from \src varies by an order of magnitude across
historical observations, the 2--10~keV flux ranging from $\sim0.6
\times 10^{-11} \ \rm erg \ cm^{-2} \ s^{-1}$ to $\sim6 \times
10^{-11} \ \rm erg \ cm^{-2} \ s^{-1}$.  The 2--10~keV flux during the
\suzaku observation is $\sim2 \times 10^{-11} \ \rm erg \ cm^{-2}
\ s^{-1}$ (\tablefluxesp), so it is in the upper end of the historical
range.  The 2--10~keV observed luminosity during the \suzaku
observation is $\sim3 \times 10^{42} \ \rm erg \ s^{-1}$.  The
intrinsic luminosity is model-dependent, and we reported the values we
obtained from modelling the \suzaku spectrum in \tablefluxesp, which
shows 2--10~keV values in the range $\sim7.4$--$8.1 \times 10^{42}
\ \rm erg \ s^{-1}$ Thus, historical observations suggest that the
intrinsic luminosity could be greater than $2 \times 10^{43} \ \rm erg
\ s^{-1}$ at times.

\begin{table}
\caption[Historical X-ray Spectral Studies of \src]{Historical X-ray Spectral Studies of \src}
\begin{center}
\begin{tabular}{ll}
\hline
Mission & References \\
\hline
\asca & Iwasawa \etal 2007 \\
\bsax & Risaliti \etal 2002 \\
\rxte & Elvis \etal 2004 \\
      & Rivers \etal 2013 \\
\integral & Beckmann \etal 2004 \\
	  & Fedorova \etal 2011 \\
	  & Ursini \etal 2019 \\
	  & Lubi\'{n}sky \etal 2016 \\
\chandra & Shu \etal 2011 \\
	 & Yi \etal 2021 \\
\xmm & Elvis \etal 2004 \\
     & Beckmann \etal 2004 \\
     & Ursini \etal 2019 \\ 
     & Lubi\'{n}sky \etal 2016 \\
\suzaku & Shirai \etal 2008 \\
	& Miyazawa \etal 2009 \\
	& Fukazawa \etal 2011 \\
	& Kawamuro \etal 2019 \\ 
	& Lubi\'{n}sky \etal 2016 \\
\swift & Fedorova \etal 2011 \\
	& Kawamuro \etal 2019 \\ 
\nustar & Masini \etal 2016 \\
	& Kamraj \etal 2017 \\
	& Ursini \etal 2019 \\ 
\nicer  & Miller \etal 2019 \\
\hline
\end{tabular}
\end{center}
\end{table}

\subsection*{Column Density} 
The column density in \src is known to be variable, and the historical
range is $N_{\rm H} \sim2$ to $7 \times 10^{23} \rm
\ cm^{-2}$. However, except for analyses of the \nustar data (Kamraj
\etal 2017; Masini \etal 2019) and the \nicer data (Miller \etal
2019), all of the other column density measurements are obtained from
simple one-dimensional models of the extinction.  Shirai \etal (2008)
reported column densities in the range $\sim2.3$ to $4.1 \times
10^{23} \rm \ cm^{-2}$ from various phenomenological fits to the
\suzaku spectrum. Miyazawa \etal (2009) and Fukazawa \etal (2011)
obtained values consistent with that range for the same \suzaku data.
We obtained a column density of $\sim2.6 \times 10^{23} \rm \ cm^{-2}$
for a uniform spherical model for the same data. For \mytorus model
fits we obtained $\sim3 \times 10^{23} \rm \ cm^{-2}$ for the
line-of-sight absorption (see \tablespfitsp).  The decoupled version
of the \mytorus model gave a global column density of $\sim6 \times
10^{23} \rm \ cm^{-2}$, By comparison, coupled \mytorus fits to the
\nustar data gave a column density of $\sim6.5 \times 10^{23} \rm
\ cm^{-2}$ (Kamraj \etal 2017), and decoupled \mytorus fits to the
\nustar data gave a column density of $\sim4.4 \times 10^{23} \rm
\ cm^{-2}$ (Masini \etal 2019).  The \nicer data gave a column density
of $\sim2.6 \times 10^{23} \rm \ cm^{-2}$ (Miller \etal 2019) for the
coupled \mytorus model.  All of these measurements are consistent with
a clumpy distribution of matter, which has an average global column
density of $\sim7 \times 10^{23} \rm \ cm^{-2}$, with clumps moving in
and out of the line of sight, resulting in a variable line-of-sight
column density.

\subsection*{Narrow \fekalfa Emission Line} 

The best measurement of the (rest-frame) centroid energy of the narrow
\fekalfa line is still that from a \chandra HETG observation (Shu
\etal 2011), its value of $6.393\pm0.004$~keV being consistent with
the weighted mean of the $K\alpha_{1}$ and $K\alpha_{2}$ components
from neutral Fe (subject to calibration systematics and velocity
shifts relative to the galaxy redshift).  All other historical
observations are consistent with this, but the different instruments
have larger systematic and statistical errors than the \chandra HETG
data. In our modelling of the narrow \fekalfa line, the energies of
the $K\alpha_{1}$ and $K\alpha_{2}$ components are fixed by physics,
but we allowed a shift to account for possible calibration systematics
and/or velocity shifts. We also concluded that the data are consistent
with neutral Fe, with a systematic, expressed as a net velocity
blueshift, of $700-900 \ \rm km \ s^{-1}$, depending on the model.

For the narrow \fekalfa line, we obtained a line flux of $\sim9.0
\times 10^{-5} \rm \ photons \ cm^{-2} \ s{-1}$ from the spherical
model, and $\sim8.7 \times 10^{-5} \rm \ photons \ cm^{-2} \ s^{-1}$
from both implementations of the \mytorus model. We obtained an
equivalent width of 257~eV from the spherical model, and 255~eV from
both implementations of the \mytorus model.  All except the \swift
study by Ferodova \etal (2011) gave explicit \fekalfa line fluxes.
The \fekalfa line flux reported for observations by other missions,
clusters in the range $\sim7$--$9 \times 10^{-5} \rm \ photons
\ cm^{-2} \ s^{-1}$, but on one occasion, in a 2011 \xmm observation,
a lower value of $5.7\pm0.8 \times 10^{-5} \rm \ photons \ cm^{-2}
\ s^{-1}$ was reported (Ursini \etal 2019). Except for this latter
observation, for the studies that do report an explicit line flux, the
values are within $\sim20\%$, or less, of our measurements from the
\suzaku spectrum. Note that fluxes across different missions are not
adjusted for absolute flux calibration systematics, which can be more
than 10 per cent in some cases.

The \fekalfa line equivalent width reported in studies based on
observations with different X-ray astronomy missions has a wide range,
from $\sim50$ eV, to $\sim700$~eV. These results are consistent with a
picture in which the \fekalfa line flux remains steady most of the
time, not responding to continuum variations, whilst the equivalent
width varies as the continuum varies. We note that the lowest values
of the equivalent width are $50\pm10$~eV and $73\pm5$~eV, from a 2011
\xmm observation (Ursini \etal 2019), and the \nicer observation
(Kamraj \etal 2017) respectively. The continuum luminosity during
these observations was at the upper end of the historical range, and
during the time of the \nicer observation, the continuum was a factor
$\sim3$ higher than it was during the \suzaku observation.

As for the narrow \fekalfa line width, the best measurement of its
width is that from a \chandra HETG observation (Shu \etal 2011),
because the HETG data have the highest spectral resolution, which is
$\sim1860 \ \rm km \ s^{-1}$ FWHM, at 6.4~keV, which is a factor of
$\sim3$ higher than CCD resolution.  The HETG line width is
$2430^{+620}_{-590} \ \rm km \ s^{-1}$ FWHM, which compares well with
our measurements from the \suzaku spectrum (the highest value in
\tablespfits is $3415^{+740}_{-590} \ \rm km \ s^{-1}$ FWHM), despite
the poorer spectral resolution. Shirai \etal (2008) gave a Gaussian
line width of $45^{+5}_{-6}$~eV for the same \suzaku data, which
corresponds to a FWHM of $4965 \ \rm km \ s^{-1}$. Their statistical
errors appear to be too small, and their larger width is due to the
fact that they used phenomenological models, whereas our widths are
based on physical models, and are closer to the \chandra HETG
measurement, which is less model-dependent than CCD measurements.
Miller \etal (2019) gave a Gaussian line width of $40\pm10$~eV from
the \nicer data, which corresponds to $4433 \ \rm km \ s^{-1}$ FWHM
(but note that the FWHM given by Miller \etal (2019) is incorrect,
because it does not correspond to their Gaussian width).  Measurements
of the \fekalfa line width from \xmm data (Elvis \etal 2004, Beckmann
\etal 2004) gave even larger values than Shirai \etal (2008), in the
region of $\sim7300$--$8500 \ \rm km \ s^{-1}$ FWHM.  All of these
large widths from CCD measurements are again due to the use of
phenomenological models. Ursini \etal (2019) do not give any line
width constraints from their \xmm analysis, nor do Fedorova \etal
(2011) from their \swift analysis. All other studies shown in
\tablehistory are based on non-CCD instruments, and have much worse
spectral resolution than CCDs, so \fekalfa line width measurements
from those data are not informative.

Shirai \etal (2008) and Kawamuro \etal (2016) included an additional
Gaussian Compton shoulder component for the \fekalfa line, and forced
its flux to be a fixed percentage of the core (unscattered) \fekalfa
line.  However, this empirical model has the wrong shape for the
Compton shoulder, and its strength relative to the unscattered line
component should depend on the column density and geometry of the
line-emitting matter. Moreover, Yaqoob and Murphy (2010) showed that
it is not possible, even in principle, to detect the Compton shoulder
with CCD detectors.  In contrast, all of the models that we applied to
the \suzaku data in the present paper self-consistently include the
Compton shoulder, so it is not an \adhoc component in our models. In
any case, the Compton shoulder makes an insignificant contribution to
the \fekalfa line profile for the column densities involved, which is
why it cannot be seen in the data or models (e.g., see Yaqoob and
Murphy 2010).

\subsection*{Reflection Continuum and Broad \fekalfa Emission Line} 
Historical X-ray spectral studies of \src (\tablehistoryp) have either
included no Compton-reflection continuum (ASCA, Iwasawa \etal 1997),
or they have fitted the nonrelativistic, neutral disk-reflection model
\pexrav (Magdziarz and Zdziarski, 1995), or \pexmon (Nandra \etal
2007).  The latter includes an associated (narrow) \fekalfa emission
line, whereas \pexrav does not.  Only the \nicer study (Miller \etal
2019) used \pexmonp, the rest used \pexravp.  Some of the studies
state that the Compton-reflection component is included to model
actual X-ray reflection from a disk, whereas other studies state that
they are using \pexrav to model Compton-reflection from the obscuring
material that is also responsible for the X-ray absorption. In the
former case, Compton-reflection from the obscuring material should
have been included (because there has to be some associated with the
neutral obscuring material that produces the \fekalfa line).  In the
latter case, the use of \pexrav as a surrogate is problematic. The
reason is that the shape of a Compton-reflection continuum depends on
the column density and the geometry, amongst other things (e.g. Murphy
\& Yaqoob 2009; Brightman and Nandra 2011), and the \fekalfa emission
line has to be added in as an \adhoc empirical component, whereas it
is physically tied to the geometry and column density of the same
material producing the Compton-reflection continuum.

The fact that \pexrav is not appropriate for modelling the
Compton-reflection continuum from the distribution of obscuring matter
has resulted in different studies making different assumptions about
the value at which the disk inclination angle in the \pexrav model
should be fixed. This inclination angle is meaningless when a disk
geometry is used as a surrogate for a geometry that is not a disk, and
different authors have fixed it at one of several values between
30$^{\circ}$ and 78$^{\circ}$. Shirai \etal (2008) even admit that the
choice of the inclination angle affects the conclusion about the value
of the so-called `reflection fraction' parameter, $R$, which is 1.0
for a disk that subtends a solid angle of $2\pi$ at the X-ray source,
which is non-variable and illuminates the disk from a point on its
axis. The inclination angle cannot be left as a free parameter,
because it would lead to the awkward problem of interpreting a
meaningless parameter in the context that the model is used. X-ray
reprocessor spectral-fitting models that self-consistently account for
absorption, and its associated Compton-reflection and \fekalfa
fluorescence emission-line, were readily available for work published
in 2011 or later, but only the \nustar (Masini \etal 2016; Kamraj
\etal 2017) and \nicer (Miller \etal 2019) studies made use of such
models (specifically, they applied the \mytorus model).  These studies
did also compare the results from fitting the \mytorus model with the
results of fitting \pexrav (\nustarp) or \pexmon (\nicerp).

Notwithstanding the caveats outlined above, the over-arching
conclusion from the historical works that included the \pexrav or
\pexmon models, is that all except Shirai \etal (2008), concluded that
these components are not strongly required by the data, and only upper
limits on the $R$ parameter were obtained. The upper limits on $R$
range from 0.09 (\nustarp, Kamraj \etal 2017) to 0.8 (one of the joint
\xmmp/\rxte observations, Elvis 2004). Note that a small, non-zero
value of $R=0.113^{+0.006}_{-0.007}$ was reported by Miller for the
\nicer data, but those data do not extend beyond $\sim10$~keV, so the
very small errors on $R$ are incongruent with the limited capability
of the data to constrain any Compton-reflection continuum (which, for
\pexrav and \pexmon peaks around $\sim20$--$30$~keV). Moreover, Miller
\etal (2019) state that the \pexmon inclination angle was fixed at
$5^{\circ}$, but the model does not allow inclination angles below
$18.2^{\circ}$.

Shirai \etal (2008) fitted various combinations of models to the
\suzaku X-ray spectrum, and obtained various values of $R$ that were
model-dependent, but all greater than unity, with non-zero lower
limits. This discrepancy compared to all the other studies is likely
to be due to several reasons.  One is that, as Shirai \etal (2008)
point out, the background-subtraction for the high-energy \suzaku data
was not yet sufficiently accurate at that time. Another is that Shirai
\etal (2008) allowed the relative normalization between the CCD
instrument (XIS) and high-energy instrument (HXD) to be a free
parameter, since cross-instrument calibration had not yet
matured. They also fixed the inclination angle of the \pexrav model at
$78^{\circ}$, a value that is higher than in any other study. There is
degeneracy in the inclination angle and $R$ parameters, such that more
edge-on values of the inclination angle require greater values of $R$
to achieve the same magnitude of the reflection continuum relative to
the direct continuum. A later work by Kawamuro \etal (2016) obtained
$R=0.08^{+0.05}_{-0.04}$ from the same \suzaku data. Note that
Miyazawa \etal (2009) and Fukazawa \etal (2011) also analyzed the same
\suzaku data, and included a \pexrav model component, but they did not
actually reveal any results for any of the fitted \pexrav parameters.

Only one of the studies in \tablehistory (Shirai \etal 2008)
explicitly tested the \src data for a relativistically-broadened
\fekalfa line, in the form of a discrete, \adhoc emission-line with a
shape based on Doppler and gravitational energy shifts. They obtained
only an upper limit on the equivalent width of $110$~eV. However,
given the \adhoc nature of the other model components, an
interpretation of this result is not clear. Miller \etal (2019) did
apply a relativistic blurring kernal model, \rdblur (Fabian \etal
1989), to their \pexmon component, but the purpose of that was to
attempt to constrain the inner extent of the {\it narrow} \fekalfa
line-emitting material. Note that Miller \etal (2019) state that the
inclination angle in the \rdblur model was fixed at $5^{\circ}$, but
the model does not allow inclination angles below $18.2^{\circ}$.
Only one of the studies (Kawamuro \etal 2016) fitted a full
relativistic disk model that includes both the reprocessed continuum
emission and broadened emission lines. The model that was fitted was
that of the ionized disk model \reflionx (Ross and Fabian 2005),
convolved with \rdblurp. However, they found that for \srcp, the model
did not improve the fit, resulting in a small upper limit on the
effective $R$ value, of 0.03.

\subsection*{Fe Abundance} 
If an Fe abundance is derived from phenomenological models consisting
of \adhoc components that are not physically self-consistent, then
that Fe abundance must be interpreted as just a parameter, and not an
actual Fe abundance. Specifically, an Fe abundance obtained from
fitting an \adhoc component for the \fekalfa emission line, combined
with a \pexrav component to model the Compton reflection continuum
from the material producing the line, is not
self-consistent. Moreover, the inclination angle in the \pexrav
component is fixed at an arbitrary value, since the model has an
inappropriate geometry.  Shirai \etal (2008) fitted these model
components to \suzaku data, leaving the Fe abundance in the \pexrav
model a free parameter. They obtained different values, all sub-solar,
for different variations on the overall model. However, these values
of the Fe abundance are not a measure of the abundance. In our fit to
the same \suzaku data using the uniform spherical model, the \fekalfa
emission line, the extinction, and the Compton reflection are all
physically self-consistent, so our value of $1.102^{+0.024}_{-0.021}$
for the Fe abundance relative to the assumed solar value, is the most
robust measurement of the Fe abundance from the X-ray spectrum of \src
to date.

\subsection*{Ionized Wind} 

In our analysis of the \suzaku spectrum of \src we found absorption
lines due to He-like and H-like Fe, indicative of a highly ionized
wind. Both of these features have been reported from a 2011 \xmm
observation (Ursini \etal 2019), and the He-like feature has been
reported by Shirai \etal (2008) from the same \suzaku data, and by
Miller \etal (2019) from the \nicer data. Although the \chandra HETG
has better spectral resolution than CCD data, neither absorption line
has been reported for existing \chandra HETG data, most likely due to
the limited statistics of these data.

\label{lastpage}


\begin{thebibliography}{}
\bibitem{} Anders E., Grevesse N., 1989, Geochimica et Cosmochimica Acta, 53, 197

\bibitem{} Arnaud K. A., 1996, in Jacoby G, Barnes J, eds., ASP Conf. Ser. Vol. 101,
Astronomical Data Analysis Software and Systems V. Astron. Soc. Pac., San Francisco, p. 17

\bibitem{} Asplund M., Grevesse N., Sauval A. J., and Scott P., 2009, ARA\&A, 47, 481

\bibitem{} Balokovi\'{c} M. et al., 2018, ApJ, 854, 42

\bibitem{} Balokovi\'{c} M. et al., 2020, ApJ, 905, 41

\bibitem{} Beckmann V., Gehrels N., Favre P., Walter R., Courvoisier T. J.-L., Petrucci P.-O., Malzac J., 2004, ApJ, 614, 641

\bibitem{} Braito V. et al., 2018, MNRAS, 479, 3592

\bibitem{} Brightman M., Nandra K., 2011, MNRAS, 413, 1206 (BN11)

\bibitem{} Dadina M., 2007, A\&A, 461, 1209

\bibitem{} Dadina M., 2008, A\&A, 485, 417 

\bibitem{} Dauser T., Garc\'{i}a J. A., Joyce A., Licklederer S., Connors R. M. T., Ingram A., Reynolds C. S., Wilms J., 
2022, MNRAS, 514, 3965

\bibitem{} Dauser T., Garc\'{i}a J., Parker M. L., Fabian A. C., Wilms J., 2014, MNRAS, 444, L100

\bibitem{} Dauser T., Garc\'{i}a J., Walton D. J., Eikmann W., Kallman T., McClintock J., Wilms J., 2016, A\&A, 590, A76

\bibitem{} Dauser T., Garc\'{i}a J., Wilms J., B\"{o}ck M., Brenneman L. W., Falanga M., Fukumura K., Reynolds C. S.,
2013, MNRAS, 430, 169

\bibitem{} Elvis M., Risaliti G., Nicastro F., Miller J. M., Fiore F., Puccetti S., 2004, ApJ, 615, L25

\bibitem{} Fabian A. C., Rees M. J., Stella L., White N. E., 1989, MNRAS, 238, 729

\bibitem{} Fedorova E. V., Beckmann V., Neronov A., Soldi S., 2011, MNRAS, 417, 1140

\bibitem{} Fukazawa Y. et al., 2011, ApJ, 727, 19

\bibitem{} Fukazawa Y., Furui S., Hayashi K., Ohno M., Higari K., Noda H., 2016, ApJ, 821, 15

\bibitem{} Furui S., Fukazawa Y., Odaka H., Kawaguchi T., Ohno M., Hayashi K., 2016, ApJ, 818, 164
  
\bibitem{} Garc\'{i}a J. et al., 2014, ApJ, 782, 76

\bibitem{} Garc\'{i}a J., Kallman T. R., Bautista M., Mendoza C., Deprince J., Palmeri P., Quintet P., 2018,
in Mendoza C., Turck-Chi\`{e}ze S., Colgan J., eds., ASP Conf. Ser. Vol. 515, Astrophysical Opacities. 
Astron. Soc. Pac., San Francisco, p. 282

\bibitem{} Heiles C., Cleary M. N., 1979, Australian J. Phys, Astrophys. Suppl., 47, 1
 
\bibitem{} Ikeda S., Awaki H., Terashima Y., 2009, ApJ, 692, 608

\bibitem{} Iwasawa K., Fabian A. C., Ueno S., Awaki H., Fukazawa Y., Matsushita K., Makishima K., 1997, MNRAS, 285, 683

\bibitem{} Kamraj N., Rivers E., Harrison F. A., Brightman M., Balokovi\'{c} M., 2017, ApJ, 843, 89

\bibitem{} Kawamuro T., Ueda Y., Tazaki F., Ricci C., Terashima Y., 2016, ApJ, 225, 14

\bibitem{} Laor A., 1991, ApJ, 376, 90

\bibitem{} Liu Y., Li X., 2014, ApJ, 787, 52

\bibitem{} Lu N. Y., Hoffman G. L., Groff T., Roos T., Lamphier T., 1993, ApJS, 88, 383

\bibitem{} Lubi\'{n}sky P. et al., 2016, MNRAS, 458, 2454

\bibitem{} Magdziarz P., Zdziarski A. A., 1995, MNRAS, 273, 837

\bibitem{} Masini A. et al., 2016, A\&A, 589, 59

\bibitem{} Miller J. M., Kammoun E., Ludlam R. M., Gendreau K., Arzoumanian Z., Cackett E., Tombesi F., 2019, ApJ, 884, 106

\bibitem{} Miyazawa T., Haba Y., Kunieda H., 2009, PASJ, 61, 1331

\bibitem{} Murphy K. D., Yaqoob T., 2009, MNRAS, 397, 1549 

\bibitem{} Nandra K., O'Neill P. M., George I. M., Reeves J. N., 2007, MNRAS, 382, 194

\bibitem{} Netzer H. 1990, in Active Galactic Nuclei,
ed. R. D. Blandford, H. Netzer, L. Woltjer (Berlin: Springer), 137

\bibitem{} Reynolds C. S., 2019, Nat. Astron., 3, 41

\bibitem{} Reynolds C. S., 2021, ARA\&A, 59, 117 

\bibitem{} Risaliti G., 2002, A\&A, 386, 379

\bibitem{} Rivers E., Markowitz A., Rothschild R., 2011, ApJS, 193, 3

\bibitem{} Ross R. R., Fabian A. C., 2005, 358, 211

\bibitem{} Saha T., Markowitz A. G., Buchner J., 2022, MNRAS, 509, 5485

\bibitem{} Shirai H. et al., 2008, PASJ, 60, S263 

\bibitem{} Shu X. W., Yaqoob T., Wang J. X., 2011, ApJ, 738 147

\bibitem{} Takahashi T. et al., 2007, PASJ, 59, 35

\bibitem{} Tanaka Y. \etal 1995, Nat, 375, 659

\bibitem{} Tanimoto A., Ueda Y., Odaka H., Kawaguchi T., Fukazawa Y., Kawamuro T., 2019, ApJ, 877, 95

\bibitem{} Thorne K. S., 1974, ApJ, 191, 507

\bibitem{} Tombesi F. \etal 2015, Nat, 519, 436

\bibitem{} Turner T. J., George I. M., Nandra K., Mushotzky R. F. 1997, ApJ, 488, 164

\bibitem{} Tzanavaris P., Yaqoob T., LaMassa S., Ptak A., Yukita M., 2021, ApJ, 922, 85

\bibitem{} Ursini F., Bassani L., Malizia A., Bazzano A., Bird A. J., Stephen J. B., Ubertini B., 2019, A\&A, 629, A54

\bibitem{} Walton D. J., Nardini E., Fabian A. C., 2013, MNRAS, 428, 2901

\bibitem{} Woo J.-H., Urry C. M., 2002, ApJ, 581, L5

\bibitem{} Wong K. C. et al., 2020, MNRAS, 498, 1420

\bibitem{} Yaqoob T., 2012, MNRAS, 423, 3360

\bibitem{} Yaqoob T., Edelson R., Weaver K. A., Warwick R. S., Mushotzky R F., Serlemitsos P. J., Holt S. S., 1995, ApJ, 453, L81

\bibitem{} Yaqoob T., Murphy K. D., 2010, MNRAS, 412, 277 

\bibitem{} Yaqoob T., Murphy K. D., 2011, MNRAS, 412, 1765 

\bibitem{} Yaqoob T., Tatum M. M., Scholtes A., Gottlieb A., Turner, T. J., 2015, MNRAS, 454, 973

\bibitem{} Yi H., Wang J., Shu X., Fabbiano G., Pappalardo C., Wang C., Yu H., 2021, ApJ, 908, 156 

\end{thebibliography}
\end{document}